\documentclass[10pt,journal,compsoc,twoside]{IEEEtran}

\ifCLASSOPTIONcompsoc
\usepackage[nocompress]{cite}
\else
\usepackage{cite}
\fi

\ifCLASSINFOpdf
\usepackage[pdftex]{graphicx}
\graphicspath{{./}{./}}
\DeclareGraphicsExtensions{.pdf,.jpg}
\else
\usepackage[dvips]{graphicx}
\DeclareGraphicsExtensions{.pdf,.jpg}
\fi

\usepackage{mathptmx}
\usepackage{times}

\usepackage{amsfonts}
\usepackage{amssymb}
\usepackage[cmex10]{amsmath}
\usepackage{array}
\usepackage{amscd}
\usepackage{pb-diagram}
\usepackage{graphicx}
\usepackage{epsf}
\usepackage{mathptmx}
\usepackage{color}
\usepackage{paralist}
\usepackage{framed}
\usepackage{wrapfig}
\usepackage{textgreek}
\usepackage{ragged2e}

\newcommand{\rd}[1]{{\color[rgb]{0.0,0.0,0.0}{#1}}}
\newcommand{\bl}[1]{{\color{black}{#1}}}

\usepackage[bookmarks,backref=true,linkcolor=black]{hyperref} 
\hypersetup{
	pdfauthor = {},
	pdftitle = {},
	pdfsubject = {},
	pdfkeywords = {},
	colorlinks=true,
	linkcolor= black,
	citecolor= black,
	pageanchor=true,
	urlcolor = black,
	plainpages = false,
	linktocpage
}

\begin{document}
	
	\title{Towards High-quality Visualization of\\ Superfluid Vortices}
	
	
	\author{Yulong Guo,
		Xiaopei Liu,
		Chi Xiong,
		Xuemiao Xu
		and Chi-Wing Fu
		
		\IEEEcompsocitemizethanks{
			\IEEEcompsocthanksitem
			Accepted for pulication by IEEE Transactions on Visualization and Computer Graphics, DOI: 10.1109/TVCG.2017.2719684.
			\IEEEcompsocthanksitem
			Corresponding author: Xiaopei Liu (liuxp@shanghaitech.edu.cn or aurorean.xp@gmail.com)
			\IEEEcompsocthanksitem
			Yulong Guo and Xiaopei Liu are with School of Information Science and Technology, ShanghaiTech University. E-mail: guoyl/liuxp@shanghaitech.edu.cn.
			
			\IEEEcompsocthanksitem
			Chi Xiong is with Institute of Advanced Studies, Nanyang Technological University, Singapore. E-mail: xiongchi@ntu.edu.sg.
			
			\IEEEcompsocthanksitem
			Xuemiao Xu is with Department of Computer Science and Engineering, South China University of Technology. E-mail: xuemx@scut.edu.cn.
			
			\IEEEcompsocthanksitem
			Chi-Wing Fu is with the Department of Computer Science and Engineering, the Chinese University of Hong Kong. E-mail: cwfu@cse.cuhk.edu.hk.
			
	}}
	
	\markboth{Accepted by IEEE Transactions on Visualization and Computer Graphics}%
	{Guo \MakeLowercase{\textit{et al.}} : Towards High-quality Visualization of Superfluid Vortices}
	
	
	\IEEEtitleabstractindextext{%
		\begin{justify}
		
		\begin{abstract}
			Superfluidity is a special state of matter exhibiting macroscopic quantum phenomena and acting like a fluid with zero viscosity.
			In such a state, superfluid vortices exist as phase singularities of the model equation with unique distributions.
			\bl{This paper presents novel techniques to aid the visual understanding of superfluid vortices based on the state-of-the-art non-linear Klein-Gordon equation, which evolves a complex scalar field, giving rise to special vortex lattice/ring structures with dynamic vortex formation, reconnection, and Kelvin waves, etc.
				By formulating a numerical model with theoretical physicists in superfluid research, we obtain high-quality superfluid flow data sets without noise-like waves, suitable for vortex visualization.
				By further exploring superfluid vortex properties, we develop a new vortex identification and visualization method: a novel mechanism with velocity circulation to overcome phase singularity and an orthogonal-plane strategy to avoid ambiguity.}
			Hence, our visualizations can help reveal various superfluid vortex structures and enable domain experts for related visual analysis, such as the steady vortex lattice/ring structures, dynamic vortex string interactions with reconnections and energy radiations, where the famous Kelvin waves and decaying vortex tangle were clearly observed. 
			These visualizations have assisted physicists to verify the superfluid model, and further explore its dynamic behavior more intuitively.
			
		\end{abstract}
		
		\end{justify}

		
		
	}
	
	

	
	
	
	
	
	
	\maketitle
		
	\IEEEdisplaynontitleabstractindextext
	

	\section{Introduction}
	
	Superfluidity is a special state of matter~\cite{Tilley-74} first discovered in liquid helium (Helium-II) in 1930s~\cite{Nature_38}.
	Since then, various superfluid phenomena have been found in different areas of physics, from condensed matter physics, high energy physics, to astrophysics.
	
	Superfluids have many surprising properties, some of which are counterintuitive.
	For example, superfluids have zero viscosity, so they can move frictionlessly; superfluids can exhibit extremely high thermal conductivity, so heat can be conducted in almost no time.
	In one particular situation, liquid helium in superfluidity can flow out of the containing cup by itself and move along the wall without any external force~\cite{Annett_2004}.
	These interesting properties and behaviors result from macroscopic quantum phenomena, which cause strong correlation among the superfluid particles.
	Therefore, superfluids behave very differently than classical fluids~\cite{Tilley-74}.
	
	Studying superfluids has several potential applications.
		For example, superfluid liquid helium has been used as a moderator to cool down neutrons.
		The Bose-Einstein condensates (BEC\footnote{BEC is a state of matter of a dilute gas of bosons cooled to temperatures very close to absolute zero.}) with weak interactions, which also sustain superfluidity, provide platforms for many studies on quantum materials and quantum information.
		Furthermore, superfluids with vortices and turbulence occur in almost all superfluid states, similar to classical vortices and turbulence in classical fluids, which are very important in many superfluid studies, with several particular properties.
	
	\vspace*{1.5mm}
	\noindent
	{\bf Superfluid modeling.} \
	Early research on superfluids relied on real experiments~\cite{Nature_38}.
	Physicists had to design complicated methods to create a specific physical environment for a matter to reach the state of superfluidity, so that they can measure the physical properties of superfluids, e.g., thermal conductivity, dynamic viscosity, etc.
	However, it is highly challenging to obtain high-precision experimental measurements~\cite{Donnelly-91}.
	Hence, it is very difficult to find out how superfluids behave subject to varying physical parameters and system configurations purely by real experiments.
	
	This motivates physicists to develop mathematical models~\cite{HXZ_14,PRD_14} to describe the dynamic behavior of superfluids.
	However, unlike classical fluids, which are governed by the well-known Navier-Stokes equations, there is no well-established superfluid model that has been thoroughly verified and is general in handling a large variety of situations.
	So far, only a few simple models~\cite{Tisza-38,Landau-41} have been experimentally affirmed, and they do not involve superfluid vortices.
	Research on superfluid modeling is still in active development.
	The research in this paper is a collaborative work with theoretical physicists who study superfluid modeling, \rd{in which} we aim to develop simulation and visualization methods to help them explore potential properties of superfluid models, as well as verify and compare their models with real observations.
	
	\begin{figure*}[t]
		\centering
		\includegraphics[width=0.98\textwidth]{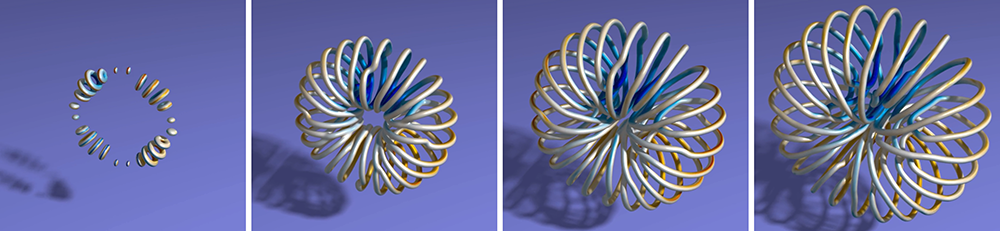}
		\caption
		{
			The formation of a steady converged 3D superfluid vortex ring produced by our simulation and visualization methods.
			The iso-surfaces are extracted based on the circulation values, and colored by velocity x-component (red for positive and blue for negative) to suggest vortex orientations.
		}
		\label{fig:teaser}
	\end{figure*}

	\vspace*{1.5mm}
	\noindent
	{\bf Superfluid vortices.}
	One distinctive characteristic of superfluids is the formation of quantum vortices~\cite{Abo-Shaeer_01,Lathrop_06}, which, in contrast to vortices in classical fluids, exhibit a quantized flux circulation of some quantity, e.g., phase.
	This leads to some special phenomena not observed in classical fluids.
	For example, a steady regularly-distributed vortex lattice can be experimentally observed in rotating superfluids, where the vortices are completely isolated from one to another.
	Superfluid vortices are basic elements in quantum turbulence.
	In addition to experimental observations in condensed matter physics, they are also applied to nuclear physics, high-energy physics, and cosmology, e.g., cosmic strings and vortex tangles in early Universe simulations~\cite{Huang-2012,HXZ_14,PRD_14}.
	
	Vortices are important features in fluids with extensive amount of work devoted to study them~\cite{PF_07,JFM_05} and to perform visual analysis~\cite{Schafhitzel_11,Koehler_11,Treib_12,Hummel_13,Clyne_13,Shafii_13} on them in classical fluids.
	Since superfluid vortices differ from their classical counterparts, mainly due to the phase singularity of the model equation, we cannot directly apply existing flow visualization methods to superfluid vortices.
	Moreover, studying superfluid vortices is an emerging research topic in physics, and much more is to be understood about their physical properties.
	Hence, we work closely with physicists to design simulation and visualization methods to explore superfluid vortices based on their expertise.

	
	\vspace*{1.5mm}
	\noindent
	{\bf Overview of this work.} \
	In this work, our goal is to {\em develop visualization methods to aid physicists to visualize and explore superfluid vortices, with an emphasis on applications to support intuitive scientific study}.
	To visualize superfluid vortices, \rd{data sets need to be first prepared typically} by means of simulations based on the governing model equations. Here in particular, we employ and solve the model equation derived in~\cite{PRD_14}, which is a very recent superfluid model, theoretically proven to be a more accurate one that can cover the entire velocity range, from zero to the speed of light.
	Note that while \cite{PRD_14} focuses on the model derivation, this work develops methods for high-quality superfluid vortex visualization, where we numerically solve the model equation with efficiency (which has not been done in~\cite{PRD_14}), enhance the quality of the resulting superfluid data sets, effectively identify and extract superfluid vortices, and then visualize different superfluid phenomenon, as well as perform associated visual study with domain experts in superfluid research.
	In summary, this work has the following key technical contributions:
	\begin{itemize}
		
		\vspace*{-0.25mm}
		\item
		First, to develop superfluid visualization, we need prepare appropriate data sets.
		As far as we know, there are no publicly-available superfluid flow data sets that could be used for visualization.
		Thus, we derive methods to generate superfluid flow data sets for visualization by numerically solving the model equation.
		To obtain a stable solution of superfluid flow in 2D, we propose transforming the underlying equation to polar coordinate domain to eliminate the numerically unstable temporal-spatial cross term, and devise a stable hybrid method.
		In 3D, rather than develop complicated numerical solver, we instead adopt model approximation and develop an efficient but stable high-order finite difference scheme.
		
		\vspace*{1mm}
		\item
		Second, to enhance the visualization quality, we devise a temporal filtering method to suppress the high frequency noise-like waves, \rd{which} significantly improves the quality of the data sets, and thus the quality of the visualizations.
		Note that this is a very important step, since \rd{without such a filtering, noise-like waves may seriously influence the vortex identification, which could lead to erroneous visualizations.}
		
		\item
		Third, due to the phase singularity of the superfluid model equation at the vortices, existing approaches are difficult to \rd{apply} for vortex visualization.
		Thus, we propose to utilize the singularity instead, and develop a new technique based on velocity circulation when identifying superfluid vortices, especially in 3D.
		In theory, we can compute such a circulation along any \rd{closed curve} around a grid point, and for simplicity, \rd{a closed curve on a plane in 3D}.
		However, \rd{ambiguities} in numerical computation of circulation may fail to find certain important vortex structures.
		Hence, we devise an orthogonal-plane strategy to obtain \rd{complete} circulation values over the field, so that we can faithfully identify a full set of superfluid vortices.
		Furthermore, together with filtering (smoothing), we can easily extract \rd{vortex tubes suitable for visualization}.
		
		\vspace*{1mm}
		\item
		{
			Fourth, as a direct application of our techniques, we perform visual study and analysis of superfluid vortices for both steady and unsteady scenarios, and present various superfluid phenomena as our visualization results.
			In the steady case, we study superfluids in a rotating frame, and explore the regularly-distributed vortex lattice and vortex ring structures.
			In addition, we examine and verify the Feynman relation~\cite{Feynman-55} in 2D in order to verify the appropriateness of the physical model.
			In the unsteady case, we study superfluids in an inertial frame, and evolve and visualize vortex strings with reconnections, Kelvin wave formation and transmission, leapfrogging, as well as vortex tangle (a turbulent state of superfluids).
			With these visualizations, scientists can obtain higher-quality visual analysis of vortex dynamics for superfluid studies as well as scientific experiments.
		}
	\end{itemize}
	
	In this work, we consider different superfluid conditions and produce assorted visualization results, either steady or unsteady; e.g., Fig.~\ref{fig:teaser} shows the dynamic formation process of a 3D superfluid vortex ring structure discovered and visualized by our method, which has never been observed before; note that the similar result presented in~\cite{PRD_14} was generated by the numerical simulation technique presented in this paper, but visualized with direct volume rendering of the field magnitude (density), which is less accurate; we will explain later that such visualization is also not generally applicable.
	Moreover, we verify and compare our method with related physical models together with domain physicists, with positive expert feedbacks.
	Furthermore, this work is the first attempt we are aware of in developing visualization method based on superfluid model simulation data sets; it is useful for visual exploration and analysis of vortex structures in superfluids, and contributes to open a new area of visualization on macroscopic quantum phenomena in fluids.

	\section{Related work} \label{sec:related work}
	
	Superfluid vortices have been observed and studied in various experiments.
	Abo-Shaeer et al.~\cite{Abo-Shaeer_01} used laser beams to confine and rotate Bose-Einstein condensates (BEC) and produced a vortex lattice structure in superfluids, while Lathrop et al.~\cite{Lathrop_06} observed turbulence with vortices in superfluid liquid helium.

	\vspace{0.15cm}
	\noindent
	\textbf{Superfluid simulation.} \
	The rotating BEC in superfluid simulations often adopts the Gross-Pitaevskii equation (GPE), \rd{which is a low-speed limit model}~\cite{Tsubota_02} often used as a tool for studying superfluid vortex dynamics. 
	Tsubota et al.~\cite{Tsubota_02} solved the GPE in 2D to simulate vortex lattice in rotating superfluids, while Kasamatsu et al.~\cite{Kasamatsu_PRA_05} solved the GPE in 3D and observed vortex tubes with lattice formation.
	Zuccher et al.~\cite{Zuccher_12} analyzed the quantum vortex reconnection, while Caplan et al.~\cite{NLSEmagic} explored vortex ring dynamics. 
	This work adopts a novel superfluid model~\cite{PRD_14}, which covers the entire velocity range and incorporates the previous models as its non-relativistic limit.
	The work of~\cite{PRD_14} focuses on deriving the equation for a more general and accurate superfluid model.
	In contrast, this work focuses instead on methods towards high-quality superfluid visualizations, including simulation and vortex identification techniques for superfluid data set preparation and vortex structure visualization.
	
	\vspace{0.2cm}
	\noindent
	\textbf{Vortex core identification.} \
	There are a number of methods for identifying vortex cores in classical fluid flow fields.
	Hunt et al.~\cite{Hunt_88} proposed the Q-criterion, while Jeong  and Hussain~\cite{Jeong_95} proposed the $\lambda_2$-criterion.
	Jiang et al.~\cite{Jiang_05} provided a taxonomy of identification methods according to multiple criteria.
	While most of the existing techniques are local, Wei\ss{}mann et al.~\cite{WeiBmann_14} formulated a global method to identify vortex core lines over a vector field based on quantum mechanics analogy.
	
	Different from the above methods, which are designed for classical fluid flows, we develop a novel method to identify superfluid vortices from a complex-valued flow field.
	Our method considers the phase singularity of vorticity in the superfluid model, and is able to effectively identify vortices in superfluid flows.


	\vspace{0.15cm}
	\noindent
	\textbf{Vortical flow visualization.} \
	There are several approaches to visualize vortices in classical fluid flows, one of which is the line-integral-convolution (LIC) visualization.
	Wiebel et al.~\cite{Wiebel_07} embedded streamlines in an LIC texture to explore boundary-induced vortices.
	Krishnan et al.~\cite{Krishnan_09} attached streamlines to flow surfaces in the visualizations.
	Yu et al.~\cite{Yu_12} created a hierarchy of streamline bundles for multi-resolution exploration of flow fields. 
	Chaudhuri et al.~\cite{Chaudhuri_14} enhanced salient features in streamline visualizations by multi-scale analysis.
	
	Feature-based methods are another common approaches in vortical flow visualizations.
	Jankun-Kelly et al.~\cite{Jankun-Kelly_TVCG_06} proposed to detect and visualize vortices in engineering environments.
	Schneider et al.~\cite{Schneider_08} used the $\lambda_2$ criterion with the largest contours to extract iso-surfaces, revealing vortical features in a turbine flow.
	Schafhitzel et al.~\cite{Schafhitzel_11} visualized hairpin vortices with iso-surface rendering.
	Treib et al.~\cite{Treib_12} developed an interactive visualization system to show extremely detailed tera-scale turbulence simulations on a desktop PC.
	
	Other vortical flow visualization methods focus on different portions and aspects of vortices in classical flows.
	Sadlo et al.~\cite{Sadlo_TVCG_06} visualized vorticity transport in an incompressible flow.
	Laney et al.~\cite{Laney_06} used Morse-Smale complex to study turbulent mixing in Rayleigh-Taylor instability.
	Weinkauf et al.~\cite{Weinkauf_07} extracted vortex cores from swirling particle motion in unsteady flows.
	Helgeland et al.~\cite{Helgeland_07} extracted and visualized vortex structures in wall-bounded turbulent flows, while Johnson et al.~\cite{Johnson_TVCG_08} proposed a framework for interactive visualization of transitional flows.
	
	Vortical flow visualization has also led to practical applications in many different disciplines.
	In medical study, Soni et al.~\cite{Soni_08} used particle trajectories to visualize flows in small bronchial tubes;
	Zachow et al.~\cite{Zachow_09} developed visualizations for exploring nasal flow;
	K\"{o}hler et al.~\cite{Kohler_13} used line predicates to extract vortices in cardiac MRI flows, and Born et al.~\cite{Born_13} also used line predicates to extract blood flow in MRI data.
	In turbine study, Shaffi et al.~\cite{Shafii_13} visualized vortices in turbulent wake flows produced in a simulated wind farm, while Koehler et al.~\cite{Koehler_11} studied the flow around the dragonfly wings.
	
	However, very little work has been done on simulating and visualizing superfluids.
	Zuccher et al.~\cite{Zuccher_12} extracted density iso-surface to visualize quantum vortex reconnection in non-relativistic superfluids.
	The method has limited accuracy and its extraction results are data-dependent. 
	Very recently, Guo et al.~\cite{Guo_16} visualized the vortices in a superconductor simulation, focusing on extracting the topology of vortex lines.
	They targeted to extract spatio-temporal vortex information, and adopted a sophisticated graph-based searching algorithm based on a global formulation.
	In contrast, \textit{our goal in this work is to visualize and analyze the spatial and temporal structures of vortices in superfluids based on our simulation data sets}, thus requiring efficient methods to support extensive physical analysis.
	
	With this in mind, we developed an efficient superfluid simulation and a vortex identification/visualization method based on the circulation field and iso-surface extraction, which is local, fast, and parallelizable.
	Using our method, vortex tubes\footnote{vortex tubes are basic vortex structure in 3D superfluid visualization} can be efficiently {\em captured} by our GPU-accelerated parallel local-examination method.
	Moreover, we formulated an efficient procedure to compute the circulation and avoid the phase jump problem; this is non-trivial to achieve if we directly perform numerical integration as in conventional methods.

	\section{Superfluid Model}
	\label{sec:superfluid_fundamentals}
	
	Unlike classical fluids, superfluids are described by a complex-valued scalar field (see $\Phi$ in Eq.~\ref{eq:nlkg}).
	Here, we first review the nonlinear Klein-Gordon (NLKG) equation \bl{derived and formulated in}~\cite{PRD_14}, which is the fundamental model equation used in our simulation and visualization.

	\vspace{0.1cm}
	\noindent
	\textbf{Nonlinear Klein-Gordon equation.}
	The reason we choose the nonlinear Klein-Gordon equation as our governing equation for superfluid flows is that such an equation can be used for the entire velocity range, especially the relativistic region in which superfluidity occurs.
	The equation comes from the quantum field theory and it has the following form:
	\begin{equation} \label{eq:nlkg}
		- \frac{\partial^2 \Phi}{\partial t^2} + \nabla^2 \Phi = f(\Phi),
	\end{equation}
	where 
	$\Phi$ is a complex scalar field defined as
	\begin{equation} \label{eq:def_phi}
		\Phi = |\Phi| e ^ {i\sigma} ,
	\end{equation}
	with magnitude $|\Phi|$ and phase $\sigma$;
	$f(\Phi)=f_n(\Phi)+f_m(\Phi)$ includes a nonlinear term $f_n(\Phi)=[ \lambda ( |\Phi|^2-F_0^2 ) ] \Phi$ and a mass term $f_m(\Phi)=m_0\Phi$ to describe the self-interaction among the superfluid particles; and
	$t$ is time.
	Note that $f(\Phi)$ is parameterized by the nonlinearity constant $\lambda$, the field magnitude at infinity $F_0$, and the field mass $m_0$.
	Here, $\Phi$ can be used to derive the hydrodynamic density $\rho_s$ and velocity $\mathbf{v}$ as: $\rho=|\Phi|^2$ and $\mathbf{v} = \nabla \sigma$,
	which will be used later in our superfluid vortex visualization.
	
	Note that in superfluids, since the flow velocity $\mathbf{v}$ results from the gradient of phase $\sigma$ ($\sigma$ is a scalar field), it is curl/vorticity-free, i.e., $\nabla\times \mathbf{v}=\nabla\times\nabla\sigma=0$ if $\sigma$ is non-singular.
	Thus, the existence of superfluid vorticity is encoded into the {\em phase singularity} of $\sigma$~\cite{Feynman-55,PRD_14}.
	The formation of superfluid vortices requires certain conditions, e.g., the rotation of the system or the interaction of the vortex rings; when superfluid vortices are formed, they never dissipate locally until they reconnect (where the dissipation comes from the radiating waves at the reconnections), and their topologies can only be changed by the reconnection of vortex lines or the decomposition of a vortex ring into smaller ones.
	
	\vspace{0.1cm}
	\noindent
	\textbf{Superfluids in rotation.} \
	The rotation of superfluids not only leads to vorticity, but also changes the Hamiltonian of the system.
	Hence, superfluid vortices spontaneously arrange into a regular lattice structure in favor of low energy under constant rotation.
	Such a non-trivial configuration is a unique characteristic of superfluids from the macroscopic quantum phenomena.
	
	To model superfluids (with vortices) in a rotating frame, we need to include a rotation term $R(\Phi)$ in Eq.~\ref{eq:nlkg} as:
	\begin{equation} \label{eq:nlkg_rotation}
		- \frac{\partial^2 \Phi}{\partial t^2} + \nabla^2 \Phi + R(\Phi) = f(\Phi).
	\end{equation}
	There are two known approaches to model $R(\Phi)$:
	
	\begin{itemize}
		
		\vspace*{1mm}
		\item{
			\textbf{Coordinate transformation formulation}~\cite{PRD_14}. \
			This approach \bl{employs coordinate transformation and} considers superfluid flows directly in a rotating frame without approximation, and formulates $R(\Phi)$ as:
			\begin{equation}
				R(\Phi) 
				= 2 (\boldsymbol{\Omega}\times\mathbf{r}\cdot\nabla)\frac{\partial \Phi}{\partial t}-(\boldsymbol{\Omega}\times\mathbf{r}\cdot\nabla)^2\Phi \ ,
				\label{eq:galilean_rotation_formula}
			\end{equation}
			where $\boldsymbol{\Omega}$ is the constant (time-invariant) angular velocity of the entire system and $\mathbf{r}$ is the distance vector to the center of rotation.
			Such a coordinate transformation is a time-dependent rotation which involves non-inertial frame of reference (the rotating frame).
			As shown in \cite{PRD_14}, the rotation term in Eq.~\ref{eq:galilean_rotation_formula} can be obtained via such a coordinate transformation.
			Although this is a physically-consistent mathematical formulation, it contains a cross term (i.e., $2 (\boldsymbol{\Omega}\times\mathbf{r}\cdot\nabla)\frac{\partial \Phi}{\partial t}$) consisting of commutative spatial and temporal derivatives of the field $\Phi$, which induces numerical instability in the simulation and requires a sophisticated numerical solver to obtain stable solutions even for steady and low-speed flows.
			
		}
		
		\vspace*{1mm}
		\item{
			\textbf{Current-current formulation}~\cite{HXZ_14}.
			This alternative approach uses virtual forces to drive the superfluids in order to produce a rotation.
			It can be considered as a model simulating local rotation systems, i.e., the angular velocity is position-dependent; \bl{Although this formulation is not as accurate as the coordinate transformation formulation, stable numerical solutions can be achieved with high computational efficiency}.
			Here $R(\Phi)$ is formulated as:
			\begin{equation}
				R(\Phi) = i\eta\rho \left(\frac{\partial \Phi}{\partial t}+\Omega\times\mathbf{r}\cdot\nabla\Phi\right) \ ,
				\label{eq:current_rotation_formula}
			\end{equation}
			where $\eta$ is a coupling constant;
			$\rho$ is a Gaussian-shaped trapping potential;
			and $\Omega$ is the angular velocity of the constant rotation.
		}
		
	\end{itemize}

	\section{Model Simulation}
	\label{sec:simulation}
	
	Since there are no publicly available solver and data sets for superfluid flows based on NLKG (with and without rotation), we \rd{first need to} numerically solve the model equation with enough accuracy to simulate the superfluid flows before we perform vortex visualization.
	To simulate high-quality superfluid flows, we devise new numerical solvers for the superfluid model equation in both 2D and 3D respectively to produce vortex structures in a relatively high resolution grid.
	
	For 2D simulations, we focus on rotating superfluids, where we adopt the coordinate transformation formulation and propose to transform the equation into polar coordinate domain, where a new hybrid numerical method based on Fourier transform and finite difference can be derived.
	Such a simulation in 2D can accurately demonstrate the essence of vortex lattice in rotating superfluids, e.g. the Feynman relation (discussed later), and can be readily implemented in the numerical computations in comparison with the 3D scenarios.
	
	For 3D simulations, we consider both rotating and free-evolving superfluids.
	The free-evolving ones, as long as initialized properly with appropriate parameters, can eventually produce the vortex tangle due to the non-linearity of the model.
	Here, we adopt the current-current formulation to reduce the numerical complexity and propose a high-order finite-difference scheme to solve the model equation in Cartesian domain.
	Note that only high-order schemes can correctly preserve the vortex shapes.

	It is worth mentioning that the NLKG equation we study is dimensionless, i.e., we enforce the speed of light $c = 1$, so that other physical quantities are scaled accordingly for more convenient and more stable numerical simulations as in many traditional fluid solvers.


	\subsection{2D Simulation}
	\label{sec:2d_simu}
	
	If we perform 2D superfluid simulations with the coordinate transformation formulation directly in Cartesian coordinate space, the cross derivatives in $R(\Phi)$ would induce strong numerical instability no matter how we tune the simulation parameters such as the time step.
	To overcome this issue, we propose to transform and compute the entire model equation in \textit{polar coordinate domain}.
	
	\vspace{0.15cm}
	\noindent
	\textbf{Polar coordinate domain formulation.}
	By transforming the model equation to polar coordinate domain, we can reduce the numerical complexity and simplify $R(\Phi)$ as:
	\begin{equation}
		R(\Phi) = 2 \Omega \frac{\partial^2 \Phi}{\partial t \partial \theta} -\Omega^2 \frac{\partial^2 \Phi}{\partial \theta^2} \ ,
		\label{eq:polar_rotation_2d}
	\end{equation}
	where $\theta$ denotes the azimuthal angle,
	and the Laplacian operator is transformed to be:
	\begin{equation}
		\nabla^2 = \frac{\partial^2}{\partial r^2} + \frac{1}{r} \frac{\partial}{\partial r} + \frac{1}{r^2} \frac{\partial^2}{\partial \theta^2} \ .
	\end{equation}
	To improve the simulation quality, we propose to decouple the cross term in Eq.~\ref{eq:polar_rotation_2d}, since it can easily cause instabilities.
	
	\vspace{0.15cm}
	\noindent
	\textbf{Polar spectral discretization.} \
	To decouple the remaining cross derivative in $R(\Phi)$, we \bl{perform Fourier transform of the scalar field $\Phi$} in polar coordinate domain \bl{along the dimension of $\theta$}:
	\begin{equation}
		\Phi(t,r,\theta) = \sum_{k} \hat{\Phi}_k(t,r) e^{ik\theta},
	\end{equation}
	and insert the expansion \bl{into Eq.~\ref{eq:nlkg_rotation} with rotation term $R(\Phi)$ formulated by Eq.~\ref{eq:polar_rotation_2d}} to obtain:
	\begin{equation} \label{nlkg_ct_fft}
		(\frac{\partial^2}{\partial t^2} - 2ik\Omega\frac{\partial}{\partial t} - \Omega^2 k^2) \hat{\Phi}_k = (\frac{\partial^2}{\partial r^2} + \frac{1}{r} \frac{\partial}{\partial r} - \frac{k^2}{r^2}) \hat{\Phi}_k - \widehat{f(\Phi)}_k \ ,
	\end{equation}
	where the hat sign ( \ $\hat{\mathrm{}}$ \ ) indicates the Fourier transform and $k$ is the wavenumber.
	It is important to note that with Fourier transform, we can separate the spatial and temporal derivatives to avoid the cross derivatives and thus enhance the stability of the simulation.
	
	\vspace{0.15cm}
	\noindent
	\textbf{Radial finite-difference discretization.}
	To perform simulations with Eq.~\ref{nlkg_ct_fft}, we further discretize the domain over radial dimension and time.
	For temporal discretization, we employ the semi-implicit Crank-Nicolson scheme\bl{\cite{Anderson_95}} with temporal averaging on the linear terms to compute the final discretization in Fourier domain.
	For spatial discretization, we approximate both the first and second radial derivatives on the right-hand-side of Eq.~\ref{nlkg_ct_fft} in polar coordinate domain by second-order central finite-difference schemes.
	The combination of the two discretizations is written as:
	\begin{eqnarray}
		& \hspace*{-24.5mm} \left(-\frac{1}{(\Delta t)^2}+\frac{ik\Omega}{\Delta t}+\frac{k^2\Omega^2}{2}+\frac{L_k}{2} \right) \hat{\Phi}_k^{n+1} \cr
		= & \left(\frac{1}{(\Delta t)^2}+\frac{ik\Omega}{\Delta t}-\frac{k^2\Omega^2}{2}-\frac{L_k}{2} \right) \hat{\Phi}_k^{n-1}
		-\frac{2}{(\Delta t)^2} \hat{\Phi}_k^n + \widehat{f(\Phi^n)}_k,
		\label{eq:2d_simu}
	\end{eqnarray}
	where $n$ is the time step index;
	$\Delta t$ is the time step size set to be 0.1;
	and the other configuration parameters are:
	domain radius $R_0$ is 120;
	$\Omega$ is 1/240;
	$\omega_0$ is 1.0;
	$\lambda$ is 6.4;
	$F_0$ is 1.0;
	$m_0$ is 5.0;
	the polar grid resolution is 256 (radial)$\times$1024 (angular);
	and $L_k(\cdot)$ is the spatial derivative operator in the radial dimension, which takes the following form:
	\begin{eqnarray} \label{eq:def_lk}
		L_k \left(\Phi_{k,j}^n\right) &=& \frac{\hat{\Phi}_{k,j+1}^n-2\hat{\Phi}_{k,j}^n+\hat{\Phi}_{k,j-1}^n}{(\Delta r)^2} 
		+ \frac{\hat{\Phi}_{k,j+1}^n-\hat{\Phi}_{k,j-1}^n}{2 r_j \Delta r} \cr
		&~& - \frac{k^2 \hat{\Phi}_{k,j}^n}{r_j^2}, \;\;\;\; j=1,2,\dots,N_r \ ,
	\end{eqnarray}
	where $\Delta r$$=$$R_0 / N_r$ is the uniform spacing along the radial dimension and $N_r$ is the number of samples along the radius.
	The grid points are cell-centered, so that $r_j = (j - 0.5) \Delta r, \;\; j=1,2,\dots,N_r$, to avoid the polar singularity like in~\cite{Mohseni_JCP_00}.
	Hence, the radial derivatives for the first grid point next to the pole does not require $\Phi_{k,j}$ across the pole, since its summed coefficient is zero according to Eq.~\ref{eq:def_lk}.
	The outer boundary along the radial dimension (not explicitly stored) is treated by a Neumann condition with finite difference discretization.
	The above treatments lead to a tridiagonal linear system, which can be solved efficiently in the simulation over time in Fourier domain.
	To obtain the final simulation result, we further apply an inverse Fourier transform.
	
	Due to the non-linearity, exact numerical stability analysis is difficult to be performed for the above derivation.
	However, the system is stable and accurate depending on the choice of $\Delta t$.
	We empirically found that a large $\Delta t$ may induce instability and sacrifice accuracy, and setting $\Delta t$$=$$0.1$ balances between efficiency and accuracy, and improves the stability in the 2D simulations.
	
	\vspace{0.15cm}
	\noindent
	\textbf{Initialization.}
	Traditionally, the study of rotating superfluids often adopts the GPE~\cite{Tsubota_02} (which is effectively a low-speed limit of the NLKG equation), which also verifies the experimental observations that rotating superfluids may converge to a vortex lattice state~\cite{Abo-Shaeer_01}.
	Such a state satisfies a stationary ansatz: $\Phi(\mathbf{x},t)=\Phi^{'}(\mathbf{x}) e^{i\omega t}$, where $\omega$ is the global temporal derivative of the phase, and $\Phi^{'}$ gives the appearance of the vortex lattice structure.
	To reach such a steady but \bl{non-stationary} state, we adopt the following setting for the initial condition:
	\begin{equation}
		\Phi_{0}(\mathbf{x}) = F_0 \rho_0 \  e^{-i N \theta} \ \ \
		\mathrm{and} \ \ \
		\Phi_{1}(\mathbf{x}) = \Phi_{0}(\mathbf{x}) \ e^{i\omega_0 \Delta t}
		\ ,
		\label {eq:def_phi_01}
	\end{equation}
	where $\Phi_{0}$ and $\Phi_{1}$ stand for the initial distributions of the first and second time steps;
	$\theta$ is the azimuthal angle of polar coordinates;
	$F_0$ is the field magnitude at infinity;
	$\rho_0$ is an arbitrary profile to introduce anisotropy and perturbations, where we choose to use a bowl shape with slight azimuthal deformation;
	$N$ is an integer, such that the initial domain is configured to have a circulation of $2 N \pi$ at the pole but nowhere else;
	this ensures the final distribution of stable vortex lattice, whose circulation is also $2 N \pi$ in total;
	$\omega_0$ specifies the global temporal derivative of the phase at the beginning; and
	$\Delta t$ is the time step.
	Note that $\omega_0$ is not a physically conserved quantity. Since the final state may not fully converge in practice, the temporal derivative of the phase is not constant in space, and $\omega$ can only be measured by spatial averaging.
	Hence, $\omega_0$, which is initially specified in Eq.~\ref{eq:def_phi_01}, may generally not equal to $\omega$ in the final state.  


	\subsection{3D Simulation}
	\label{sec:3d_simu}
	
	We may extend Section~\ref{sec:2d_simu} to 3D simulations of superfluids in rotation by using spherical coordinates and adopting spherical harmonics.
	However, this may complicate the solution and introduce excessive computational overhead.
	Hence, we make a compromise between accuracy and computational efficiency and adopt a model approximation approach where the current-current formulation is used in 3D simulations with rotation.
	In the following, we consider two different scenarios of 3D superfluid simulations:

	
	\vspace{0.15cm}
	\noindent
	\textbf{Scenario 1: 3D rotating superfluids.} \
	Since the current-current formulation does not couple the spatial and temporal derivatives, we can efficiently discretize the equation.
	In detail, we handle the first and second derivatives of both the spatial and temporal terms in Cartesian coordinates by second and fourth order central finite-difference schemes in time and space, respectively, in explicit form and with conditional stability.
	Note that high-order spatial discretization should be used to reduce the discretization anisotropy, which may distort the shape of the vortex lattice.
	Hence, the discretization leads to the following linear system:
	\begin{eqnarray}
		\left( - \frac{1}{(\Delta t)^2} -\frac{i\eta\rho}{2\Delta t} \right) \Phi^{n+1} 
		= \left( \frac{1}{(\Delta t)^2} -\frac{i\eta\rho}{2\Delta t} \right) \Phi^{n-1} \cr
		- L(\Phi^n)  - \frac{2}{(\Delta t)^2} \Phi^n + R^s(\Phi^n) + f(\Phi^n) \ ,
	\end{eqnarray}
	where $R^s(\Phi)$ is the spatial component of $R(\Phi)$ in the current-current formulation: $R^s(\Phi) = i \eta \rho \Omega (-y\partial_x + x\partial_y) \Phi$ and $L(\cdot)$ is the discrete Laplacian operator.
	In our formulation, we introduce a global rotation about the z-axis, so $R^s(\Phi)$ has no z-component.
	Moreover, since the linear system is diagonal, we can solve it efficiently with the following parameters: the domain side length $l=8.0$; $\Omega=90.0$; $\eta=1.0$; $\lambda=1.0$; $F_0=1.0$; $m_0=0.0$; and the grid resolution is $512^3$.
	The system is conditionally stable depending on the choice of $\Delta t$, and large $\Delta t$ reduce simulation accuracy and increase system instability.
	In our simulations, we set $\Delta t$ to be 0.004 to ensure both stability and accuracy.
	The periodic boundary condition is applied at the domain boundary and arbitrary initialization can be used inside the domain provided that there is an initial perturbation with sufficient energy on $\Phi$.
	
\begin{figure}[t]
		\centering
		\includegraphics[width=\columnwidth]{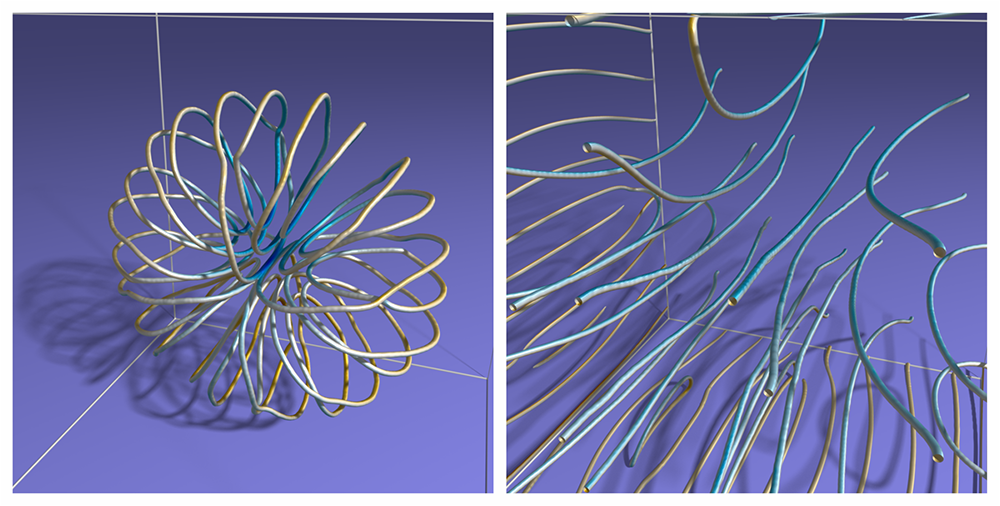}
		\caption{By adjusting the system parameters in the superfluid simulation, rotating superfluids in 3D may converge to two different states: a closed vortex ring state (left) and a vortex line state (right).}
		\label{fig:3d_rotation}
	\end{figure}

	By adjusting the trapping potential $\rho$, we can obtain different vortex lattice solutions.
	For example, by using a steep Gaussian-shaped trapping potential (which is a 3D Gaussian with smaller deviation and larger magnitude), we can obtain the vortex ring lattice shown in Fig.~\ref{fig:3d_rotation} (left).
	However, by using an oblate Gaussian-shaped trapping potential (which is a more flat 3D Gaussian with larger deviation and smaller magnitude), some straight vortex lines could emerge in the solution, see Fig.~\ref{fig:3d_rotation} (right).
	
	\vspace{0.15cm}
	\noindent
	\textbf{Scenario 2: 3D superfluid vortex dynamics.} \
	On the other hand, we can stop the rotation by giving zero angular velocity and create 3D simulations to show how vortices dynamically evolve and interact over time without a global rotation.
	In this case, since vortices do not occur spontaneously, we need to prepare some initial vortices in the simulation domain.
	
	Our first approach to prepare initial vortices is to extract the vortices produced from Scenario 1 and take them to create the initial condition before evolution.
	Specifically, we take a volume of data from the previous simulation result as $\Phi_0$, and set $\Phi_1$ according to Eq.~\ref{eq:def_phi_01}, where $\omega_0$ is tweaked to be compatible with other parameters.
	The asymmetry in the initial condition results in complex vortex motion, such as vortex reconnection and Kelvin waves.
	However, this initial condition is restricted by the number of vortex lines generated from the 3D rotating superfluids, which is relatively sparse. 
	Thus, it cannot produce sufficiently dense vortex interaction dynamics, and then vortex tangle.
	Moreover, the vortex tangle always decays during interaction, meaning that we will have fewer and fewer vortices as time proceeds.

	To generate more interesting and more complex vortex dynamics with much denser vortex tangle, 
	we employ a boundary injection approach, which is to artificially ``shoot'' closed vortex rings from the boundary of the simulation domain.
	This is physically feasible to do since we can produce the vortex rings easily with simple mechanism similar to producing a smoke ring.
	To do so, we first prepare a small volume of $\Phi$ containing a single closed ring.
	Such a small volume of closed ring can be directly shifted to the boundary edge of the whole simulation domain due to the periodic boundary setting. 
	The ring can be added to the simulation by multiplying the prepared field volume with $\Phi^{i-1}$ and $\Phi^{i}$ at some randomly selected time step $i$. 
	We then continuously place randomly shifted rings at the boundary to simulate rings coming from the boundary. 
	Finally, we stop placing rings when the underlying vortex dynamics produces sufficiently dense tangles, and we let them evolve freely to generate chaotic vortex structures.
	
	\begin{figure}[!t]
		\centering
		\includegraphics[width=0.95\columnwidth]{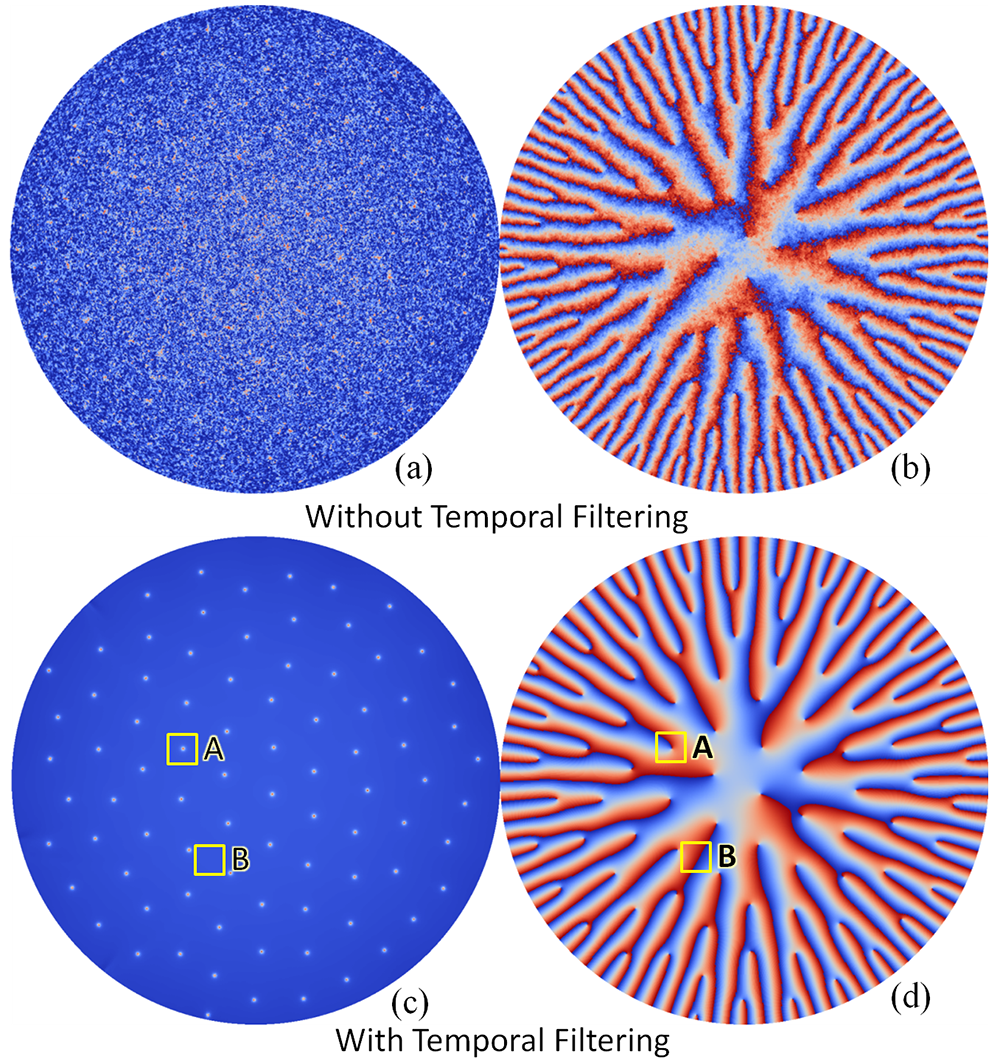}
		\caption{Comparing simulations of 2D rotating superfluids with and without temporal filtering.
			Left and right columns show the visualizations of $|\Phi|$ and phase $\sigma$, respectively; red to blue colors indicate small to large values.
			Note that the dots in (a) \& (c) are isolated vortices due to macroscopic quantum phenomena; these vortices coincide with the singularities (branch points) in the phase visualizations, e.g., branch point \textbf{A} corresponds to a vortex but not for non-branch point \textbf{B}.}
		\label{fig:2d_filter_comparison}
	\end{figure}


	\subsection{Temporal Filtering}
	\label{sec:tempral_filtering}
	
	Due to non-linear self-interaction among superfluid particles, very high-frequency waves could appear in the simulation results; this situation is physically intrinsic but not numerical due to the compressibility nature of the model equation.
	Since different bands of these waves interfere with one another in different directions, the composed waves look like noise in spatial domain, see Fig.~\ref{fig:2d_filter_comparison} (top), which significantly deteriorate the visual inspection of vortices.
	In principle, these high-frequency waves are plausible solutions to the NLKG equation.
	However, they only act as small-scale compression waves and are of little interest to the domain scientists for studying superfluid vortices, since superfluid vortices are much more structured and stationary with much lower temporal frequency than these noise-like waves.
	To get a clearer physical picture, the high-frequency waves must be removed during the simulation.
	To address this issue, we may introduce a damping term into the model equation (Eq.~\ref{eq:nlkg}) as has been applied in Bose-Einstein condensates simulations~\cite{Tsubota_02}, but this method will result in excessive energy loss and hinder the formation of a vortex lattice, which requires finite $\omega$ in the final steady state.
	
	From preliminary simulation results, we observe that such high-frequency waves are temporally incoherent even though they vary significantly over time; in addition, compared to the variation of vortices, they change much faster over time.
	Hence, we adopt a temporal filtering method to suppress the influence of these noise-like waves.
	In detail, we construct a temporal low-pass filter~\cite{OSB_99} in both 2D and 3D simulations, and apply it to the magnitude of $\Phi$ only.
	Mathematically, the filter is expressed as:
	\begin{equation}
		\widetilde{\Phi}^{n+1} = (\alpha |\Phi^{n+1}| + (1 - \alpha) |\Phi^n|) \frac{\Phi^{n+1}}{|\Phi^{n+1}|} ,
	\end{equation}
	where $\alpha\in[0,1]$ controls the filter strength, typically chosen to be close to 1.0.
	With this temporal filtering, we can significantly enhance the quality of the simulation results (see Fig.~\ref{fig:2d_filter_comparison} (bottom)) for better vortex core identification and clear vortex structure visualization. 
	This filtering technique differs from the traditional filtering methods, since we isolate magnitude and phase as independent components, where it is nontrivial to conduct spectral analysis.
	However, this technique has proven to be very effective since the dynamics is described by a complex-valued equation, where the magnitude and phase may have independent modes. 
	Furthermore, it is easy to see that our technique does not influence the stationary solution, where the magnitude has only the lowest mode with the phase being a single finite mode.

	\section{Vortex Identification and Visualization}
	\label{sec:vortex_tube_extraction}
	
	After simulating and obtaining the superfluid flow, we can identify vortex cores from the data set in order to extract the vortex structures for visualization.
	Note that we aim to visualize the structures for vortex cores only, but not the entire structure of vortices.
	Thus, we ignore other structures such as the spiral patterns, which are fluctuating structures around the vortex cores that occur at very small scales.
	To overcome the difficulty that superfluid vorticity is singular, our key idea is to employ velocity circulation and develop a specific circulation-based numerical method with an orthogonal-plane strategy to efficiently identify vortex cores while avoiding ambiguity in the circulation computation with accuracy up to the grid resolution.
	Here, we want to emphasize that sub-grid scale vortices cannot be faithfully differentiated without ambiguity; we will discuss in more detail later.


	\subsection{Observation: Circulation in Superfluids}
	As stated in Eq.~\ref{eq:def_phi}, the superfluid flow field $\Phi$ is denoted as $|\Phi| e^{i\sigma}$, where $\sigma$ (phase) is singular at the vortex cores.
	It is known from both existing theory and our visualization (see Fig.~\ref{fig:2d_filter_comparison}(c)\&(d)) that superfluid vortex cores coincide with the singularity in phase space, where the phase value is undefined, e.g., the branch point \textbf{A} in Fig.~\ref{fig:2d_filter_comparison}(d).
	This is a unique characteristic of the superfluid model different from the classical model, where vortices do not possess singularities for the related physical quantities over the flow field.
	In addition, superfluids are potential (irrotational) flows except at the vortex cores, where phase singularities cause the vortices.
	
	From the above analysis, it is easy to think of circulation, which can be employed for vortex core identification.
	If we compute the circulation along a small closed loop (either 2D or 3D) around a singular point in the phase field, e.g., a closed loop around point \textbf{A} in Fig.~\ref{fig:2d_filter_comparison}(d), we will find that such circulation is an integer multiple of $2\pi$; the integer number indicates the number of vortex cores inside the closed loop.
	However, if we compute the circulation around any non-singular point, e.g., point \textbf{B} in Fig.~\ref{fig:2d_filter_comparison}(d), even if the phase is discontinuous, we will find that the circulation is always zero.
	This is the key criterion that we will explore to formulate our vortex core identification method, with which we will employ to visualize and analyze the vortices.


	\subsection{Vortex Core Identification}
	In general, the circulation $C$ is computed as an integral:
	\begin{equation} \label{circ_def}  
		C=\oint_\mathrm{L} \mathbf{v} \cdot \mathrm{d} \mathbf{l} 
		= \oint_\mathrm{L} \nabla \sigma \cdot \mathrm{d} \mathbf{l}
		= \oint_\mathrm{L} \mathrm{d} \sigma \ ,
	\end{equation}
	where $\textbf{v}$ is the flow velocity, $\sigma$ is the phase of the complex-valued scalar field $\Phi$, and $\mathrm{L}$ is a closed loop in the phase field.
	Since superfluid vortices are usually very small with infinitesimally small vortex cores, we use a small loop (could be very arbitrary) to enclose a point, so that the resulting circulation is always $2\pi$ near the vortices, and zero elsewhere.
	Hence, we can produce a well-defined circulation field, and then threshold it to identify vortices.
	Here, we use a threshold value $\epsilon=\pi$, above which the points are considered as vortices; such a value can allow certain tolerance of circulation fluctuation from numerical evaluation with very coarse grid points.
	However, $\epsilon$ can be fine-tuned manually, but should help to successfully separate vortex points from non-vortex points.
	
	\begin{figure}[!t]
		\centering
		\includegraphics[width=0.98\columnwidth]{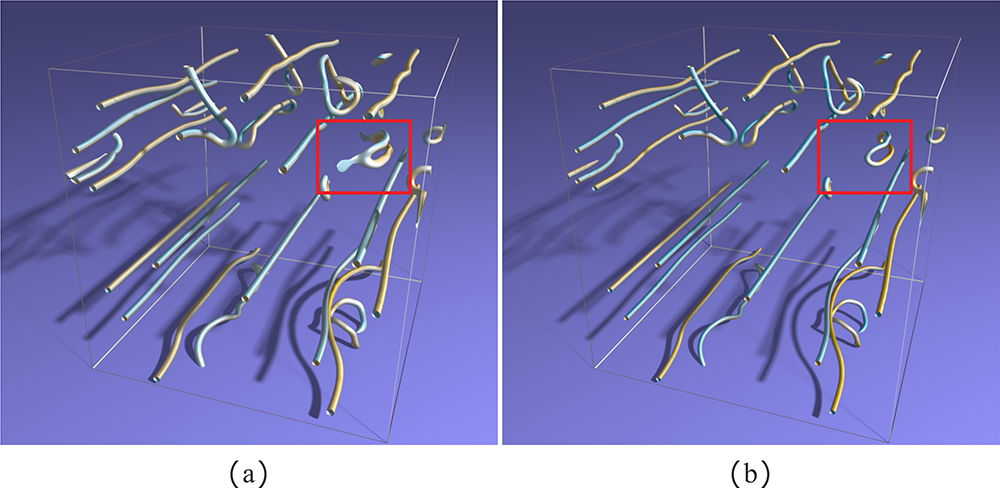}
		\caption{Comparing vortex identification methods: (a) by extracting iso-contour from the density field, and (b) our proposed method based on the circulation field.
			It can be seen from the red box that some incorrect vortex structures are identified by the density iso-contour method.}
		\label{fig:3d_density_circulation_compare}
	\end{figure}
	
	It is worthwhile to note that existing works on identifying quantum vortices usually rely on density iso-contour~\cite{Zuccher_12} (equivalently the magnitude of $\Phi$), \rd{which identifies vortices simply by thresholding the density value, and it drops to zero near vortex cores but being flat for other regions.}
	However, when vortices undergo reconnections, not only their shapes change topologically, but also energies radiate out, see the red boxes in Fig.~\ref{fig:3d_density_circulation_compare}, which show a 3D simulation of free-evolving superfluid vortex dynamics after a reconnection.
	Such energy radiation significantly affects the superfluid density, so the density iso-contour approach may generate unwanted non-vortex structures and false positive superfluid vortex structures, see Fig.~\ref{fig:3d_density_circulation_compare} (a).
	In Fig.~\ref{fig:3d_density_circulation_compare} (b), our vortex identification method, which makes use of the circulation property, can correctly capture the vortices without producing additional non-vortex structures, see the corresponding red box.
	
	As mentioned in Section~\ref{sec:related work}, Guo et al.~\cite{Guo_16} proposed a comprehensive approach to identify superconductor vortices by detecting phase-singularity-induced circulation.
	We also adopt the circulation to identify superfluid (quantum) vortices.
	However, we propose a more practical framework, which is local and fully parallelizable, enabling efficient computation on the GPU.
	

	\subsection{Numerical Evaluation}
	
	To compute a circulation field, one may arrange a dense rectangular grid over the spatial domain and compute the circulation integral (Eq.~\ref{circ_def}) along a small loop around each grid point.
	However, since $\sigma$ is periodic (range: $[0,2\pi)$), discontinuities appear across the period of $2\pi$, resulting in unstable evaluation of $\mathrm{d} \sigma$ and leading to erroneous numerical integration.
	To overcome this issue, we first note that for a complex-valued field $\Phi$, its phase is defined as:
	\begin{equation}
		\sigma = \arctan \left( \frac{\Im(\Phi)}{\Re(\Phi)} \right),
	\end{equation}
	\noindent
	where $\Re(\Phi)$ and $\Im(\Phi)$ are the real and imaginary parts of the field, respectively.
	If we denote $\mathit{R}=\Re\left[\Phi(\mathbf{x})\right]$ and $\mathit{I}=\Im\left[\Phi(\mathbf{x})\right]$ at location $\mathbf{x}$, we can reformulate \rd{$C$} in Eq.~\ref{circ_def} as:
	\begin{equation} \label{sigma_transform}
		C=\oint_\mathrm{L} \mathrm{d} \sigma = \oint_\mathrm{L} \mathrm{\frac{\mathit{R} d \mathit{I} - \mathit{I} d \mathit{R}}{\mathit{R}^2+\mathit{I}^2}} = \oint_\mathrm{L} \mathrm{\frac{\mathit{R} d \mathit{I} - \mathit{I} d \mathit{R}}{|\Phi|^2}} \ ,
	\end{equation}
	which can be used for consistent numerical integration since both $\mathit{R}$ and $\mathit{I}$ are continuous.
	In practice, the field magnitude variation around a vortex may be very large, and the integrand may change steeply near the singularity due to division by $|\Phi|^2$.
	It is obvious that direct numerical integration using the trapezoidal rule requires smooth variation of the integrand to preserve accuracy. 
	When sampling $\Phi$ at locations very close to the vortex cores, \rd{the integral becomes inaccurate due to large variations}. 
	On the other hand, \rd{the circulation is used only to indicate singularity}, which is independent of the field magnitude.
	Therefore, we normalize the field \rd{$\Phi$} to unitary magnitude before computing the circulation, and reformulate the circulation as the integral of \rd{the gradient of the normalized phase field}:
	\begin{equation}
		C'=\oint_\mathrm{L} \mathrm{\mathit{R}' d \mathit{I} - \mathit{I}' d \mathit{R}} \ ,
	\end{equation}
	where $\mathit{R}'$ and $\mathit{I}'$ are the real and imaginary parts of the normalized $\Phi$ field.
	We will discretize $C'$ in our practical computations.
	
	\begin{figure}
		\centering
		\includegraphics[width=0.9\columnwidth]{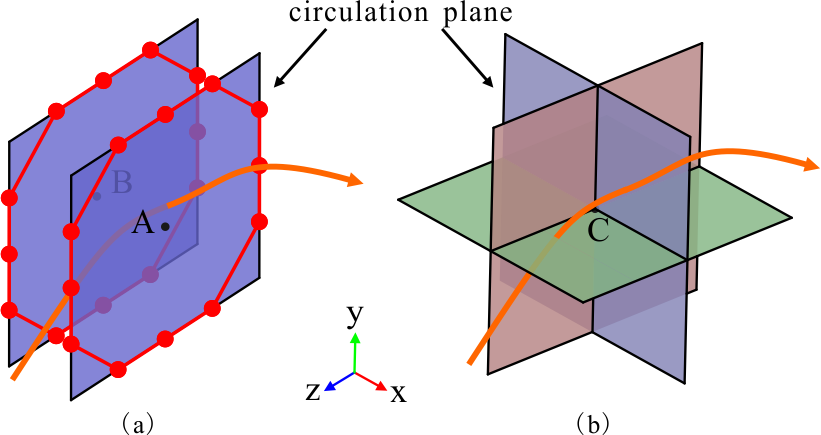}
		\caption{Circulation ambiguity in computing the circulation.
			(a) circulation detection planes at grid points are all parallel in a specific orientation; some vortices will be missed out due to circulation ambiguity; and
			(b) three orthogonal detection planes at each grid point to avoid circulation ambiguity.
			Note that the integral loops in (a) are marked in red.}
		\label{fig:circulation_ambiguity}
	\end{figure}
	
	\vspace{0.15cm}
	\noindent
	\textbf{Closed loop selection.} \
	In principle, the closed loop $\mathrm{L}$ can be selected arbitrarily (note that in 2D, the loop is a closed planar curve, while in 3D, the loop is a closed 3D curve), but arbitrary loops could unnecessarily introduce additional computation and inaccuracy due to data interpolation.
	Therefore, we align loops with the simulation grid, such that we can simply and exactly fetch the data from the neighborhood grid points.
	In 3D, we form loops on axis-aligned planes (either x-, y-, or z-plane) to construct a circulation loop at each grid point.
	It is worth to note that when using a small loop, the integral may not be sufficiently accurate due to large variation of the data near the vortex core.
	However, when using a large loop, more than one vortex core may be included in the loop and will be identified as a single vortex core, which cannot reflect the true structure of the vortices up to the precision of grid resolution.
	As a compromise, we use a small loop confined to an axis-aligned plane, which contains the nearest neighbor of the grid point on that plane as the integral loop, see Fig.~\ref{fig:circulation_ambiguity}(a), but we discard the four corners to make the loop closer to a circle in order to avoid artifacts that the identified vortex will have a box-shaped cross-section due to a box-like path.
	By dividing the path L into $N$ segments, the numerical circulation can be re-expressed as:
	\begin{equation} \label{numeric_circ}
		C' = \oint_\mathrm{L} \mathrm{\mathit{R}' d \mathit{I} - \mathit{I}' d \mathit{R}} \approx \sum_{i=1}^N \int_{\Delta \mathrm{L}_i} \mathit{R' \mathrm{d}I' - I' \mathrm{d}R'},
	\end{equation}
	where $\Delta \mathrm{L}_i$ is the $i$-th segment between two consecutive sample points,
	and the corresponding integral can be easily approximated numerically by the following trapezoidal rule:
	\begin{eqnarray}
		\int_{\Delta \mathrm{L}_i} \mathit{R' \mathrm{d}I' - I' \mathrm{d}R'} &=& 
		\rm
		\frac{\mathit{R'}(\mathbf{x}_i^s)+\mathit{R'}(\mathbf{x}_i^e)}{2} [\mathit{I'}(\mathbf{x}_i^e) - \mathit{I'}(\mathbf{x}_i^s)]  \cr
		\rm
		&-&  \frac{\mathit{I'}(\mathbf{x}_i^s)+\mathit{I'}(\mathbf{x}_i^e)}{2} [\mathit{R'}(\mathbf{x}_i^e) - \mathit{R'}(\mathbf{x}_i^s)] \cr
		\rm
		&=& \mathit{R'}(\mathbf{x}_i^s) \mathit{I'}(\mathbf{x}_i^e) - \mathit{R'}(\mathbf{x}_i^e) \mathit{I'}(\mathbf{x}_i^s),
	\end{eqnarray}
	where $\mathbf{x}_i^s$ and $\mathbf{x}_i^e$ are the locations of the starting and end points of the $i$-th segment, respectively.
	
	Note that since we use a loop with finite cross-section, we do not expect to locate the vortex cores exactly since they are infinitesimally small.
	Instead, we confine the vortex core within a finite tube with approximately twice the size of the grid spacing.
	Although we are not sure exactly the location of the vortex core, we are certain that the cross-section includes vortex cores whose mutual distances are smaller than the radius of the cross-section, implying that such a circulation-based scheme involves the aggregate of all small-scale vortices, and the shape of the tube indicates the spatial structure of the vortex cores; such spatial structure can be determined by the grid resolution.
	For vortices in sub-grid scale, uncertainties may still occur, which can only be rectified by increasing the grid resolution.
	
\begin{figure}[!t]
		\centering
		\includegraphics[width=0.98\columnwidth]{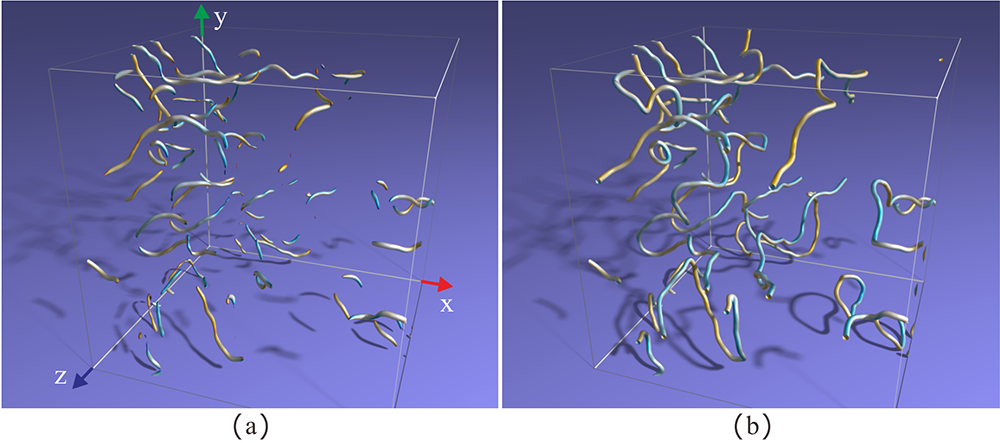}
		\caption{Comparison of vortex detection results:
			(a) detection by planes in the x-orientation, resulting in a lot of (falsely) broken vortex tubes; and 
			(b) vortex detection result by the orthogonal-planes strategy.}
		\label{fig:3d_plane_selection_comparison}
	\end{figure}

	\vspace{0.15cm}
	\noindent
	\textbf{Circulation ambiguity.} \
	For 2D simulations, the circulation field constructed above can effectively identify the vortex cores.
	However, for 3D simulations, if we compute the circulation field on planes all with one specific orientation, e.g., all on y-z planes, we may miss some core vortex structures, see Figs.~\ref{fig:circulation_ambiguity}(a) and \ref{fig:3d_plane_selection_comparison}(a).
	The reason here is that the vortex core lines, which are parallel to the circulation loop planes, may not be effectively identified by the loops on those parallel planes.
	
	In general, a vortex core line may or may not pass through a local circulation loop plane depending on the orientation of the plane.
	Since vortex cores may not overlap with the grid points, the numerical evaluation of circulations on planes almost parallel to the orientation of the vortex core lines will give close-to-zero value; hence, the detection of a portion of the vortex core lines may fail, and the grid point may be mis-identified as a non-vortex point.
	Fig.~\ref{fig:circulation_ambiguity} (a) shows the vortex core line in orange (arrow) and two parallel circulation planes at grid points \textbf{A} and \textbf{B}.
	At \textbf{A}, the vortex core line passes through the loop on the circulation plane, so we can successfully identify the vortex core.
	However, at the nearby point \textbf{B}, since the local vortex core line is parallel to the circulation plane, the vortex core identification fails even though the figure suggests the existence of a vortex core at \textbf{B}.
	Such ambiguous case becomes indistinguishable from the real case of a non-vortex point.

	\begin{figure}
		\centering
		\includegraphics[width=\columnwidth]{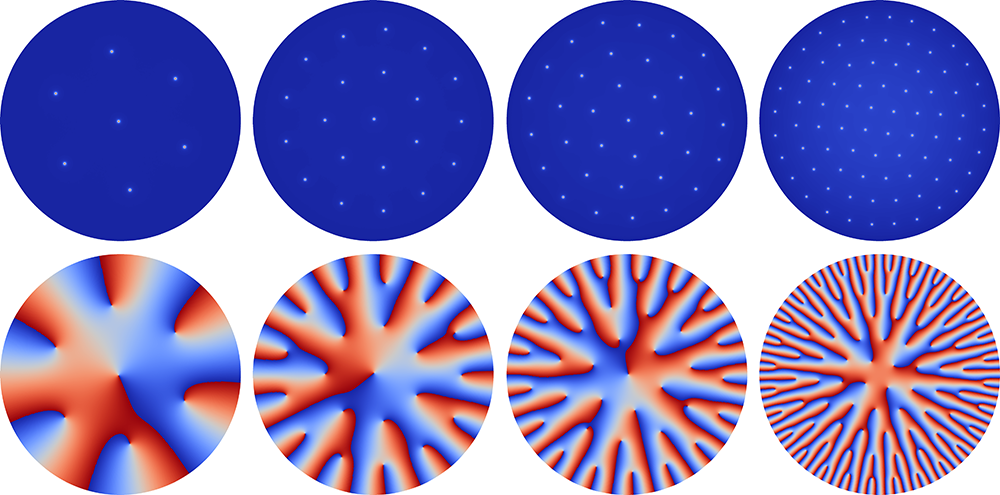}
		\caption{2D vortex lattices simulated with increasing $\omega \Omega$ (from left to right).
			Top row: magnitude of $\Phi$.
			Bottom row: phase $\sigma$.
			Here, we use the same color coding scheme (blue and red) as in Fig.~\ref{fig:2d_filter_comparison}.}
		\label{fig:2d_lattice_result}
	\end{figure}

	\vspace{0.15cm}
	\noindent
	\textbf{Orthogonal-plane strategy.} \
	To resolve such ambiguity issue, we propose to use a set of three orthogonal planes at each grid point and compute the circulation values independently on each plane, see Fig.~\ref{fig:circulation_ambiguity}(b).
	Note that these three orthogonal planes form a complete set of basis planes that can be used to construct any planes that pass through a grid point for circulation evaluation.
	This means that vortex core lines that pass through the vicinity of a grid point with arbitrary orientation can be effectively captured by the circulation values from at least one of the three evaluations.
	
	Thus, taking the maximum of the three circulation values (one for each grid planes) as our final circulation value can effectively and stably identify all vortex cores at the precision of given grid resolution without the circulation ambiguity issue described in the previous sub-section.
	Fig.~\ref{fig:3d_plane_selection_comparison} compares the vortex core identification results, where in Fig.~\ref{fig:3d_plane_selection_comparison}(b), a complete vortex identification results can be obtained using our proposed strategy.
	Also note that such a strategy, which is local in computation, can be efficiently parallelized by our GPU implementation.
	

	\subsection{Vortex Visualization}
	After vortex identification, we obtain a binary field to indicate \rd{whether each grid point is a vortex point or not}, and we perform vortex visualization based on such a field.
	Note that since we compute circulation with a loop size of two grid cells for vortex identification, such a binary field only indicates whether there are vortex cores \rd{inside the loop}, and there is almost no chance for a grid point to be exactly the vortex core.
	On the other hand, vortex cores are infinitesimally small, but visualizing them requires structures with finite size, usually using a tube structure.
	Thus, locating exactly the vortex core position is not necessary for visualization, and the binary field already forms a good candidate set for vortex tube generation， which can be used for visualization based on iso-surface extraction and rendering.
	
	However, if we directly use this binary field to generate the vortex tube, severe spatial aliasing artifacts (note that such aliasing is due to insufficient spatial sampling and is different from those noise-like waves in Section~\ref{sec:tempral_filtering}) would occur because the binary field is like a rectangle function which is discontinuous across the edges.
	To remove aliasing, we apply a Laplacian-diffusion filter with several iterations to smooth the binary field, which effectively removes the high frequency components beyond the sampling rate.
	Hence, we can generate the vortex tubes by extracting iso-surfaces from such a smoothed field.
	In our visualizations, we construct the vortex tube iso-surfaces based on an iso-value (ranged $[0,1]$) that relates to the radius of the vortex tube, and we set it as $0.5$, but it can be varied to make the tube thicker or thinner, but no thicker than several cells depending on the degree of smoothing.
	
	Note that the purpose of such a smoothing is to remove aliasing artifacts arisen from the sampling of high frequency signals.
	Since the smoothing filter is isotropic, the vortex tube generated from the smoothed field still represents the vortex core structure.
	However, smoothing may introduce uncertainty to visualization, e.g., very close vortex tubes may be merged into one vortex tube. However, these uncertainties are restricted in scales which are roughly smaller than the size of the smoothing kernel.
	To maintain faithful vortex structure, the above smoothing cannot be excessive.
	To control the degree of smoothing, we control the kernel size of the Laplacian-diffusion filter as well as the number of iterations, which can be fine-tuned empirically.

	\section{Visual Analysis}
	\label{sec:visual_analysis}
	
	\bl{Once we identify the vortex cores and extract the vortex tubes around}, we can \bl{render the vortex tubes in order to} visualize the superfluid vortex structures and perform visual analysis \rd{to answer} some important questions with our physicist collaborators.
	In addition, this research work also serves to help our collaborators explore various interesting phenomena and properties of superfluid vortices.
	In summary, our collaborators are particularly interested in the following \rd{questions} on superfluids:
		\begin{itemize}
			\item{Whether the underlying superfluid model conforms to the real experiments and describes the true physics.}
			\item{How the superfluid vortices distribute when they are under constant rotation, in both 2D and 3D.}
			\item{Whether the Feynman relation is satisfied.}
			\item{How superfluid vortices reconnect to form new vortices.}
			\item{How waves (including the helical Kelvin waves) are generated and transmitted through vortex-vortex interactions.}
			\item{How vortex tangles are formed and how they behave, etc.}
		\end{itemize}
	These questions cannot be directly answered by traditional analysis with the original model equation, even with the help of numerical simulations.
	In the following, we will visualize superfluid vortices and analyze their properties for these questions.
	\textit{Note that the supplementary video presents animations for the resulting vortex structures in superfluids, for both steady and unsteady (dynamic) scenarios}.


	\subsection{Steady Vortex Dynamics}
	\label{sec:steady_vortex_dyanmics}
	
	For superfluids in constant rotation, steady states can be reached with stable vortex distributions with appropriate parameter settings.
	Hence, we can study both 2D and 3D steady vortex structures in the presence of vortex dynamics.
	
	\vspace{0.15cm}
	\noindent
	\textbf{Vortex lattice formation in 2D.} \
	The 2D simulation of superfluids in constant rotation can be directly visualized using the scalar field magnitude.
	Through the visualization, we can then observe regularly-distributed vortex lattice structures, where the vortices are completely isolated and located repulsively to maintain low energy, see Fig.~\ref{fig:2d_lattice_result}.
	In detail, we maintain the physical validity of the simulation by imposing a constraint $\Omega R_0 <1$, so that the rotation does not violate the limit of the speed of light.
	Moreover, it is also important to ensure a sufficiently large $\omega_0 \Omega$ in the initial condition, so that we can maintain the vortex lattice when the field converges to a finite $\omega \Omega$.
	
	\begin{figure}[!t]
		\centering
		\includegraphics[width=0.8\columnwidth]{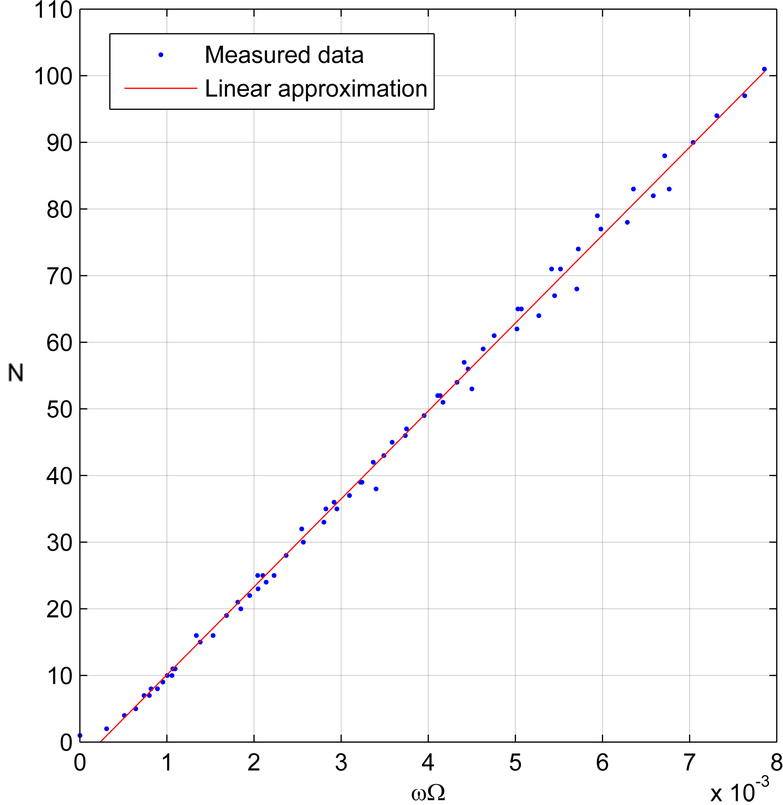}
		\caption{Verification of the Feynman relation by 2D vortex lattice simulation: by automatically detecting vortex cores and counting their numbers ($N$) with respect to different angular velocities ($\omega \Omega$), we can plot the samples and fit a (linear) line \bl{in the least square sense}, showing that our results closely agree with the Feynman relation~\cite{Feynman-55}.}
		\label{fig:feynman_relation}
	\end{figure}
	
	\begin{figure}[!t]
		\centering
		\includegraphics[width=0.98\columnwidth]{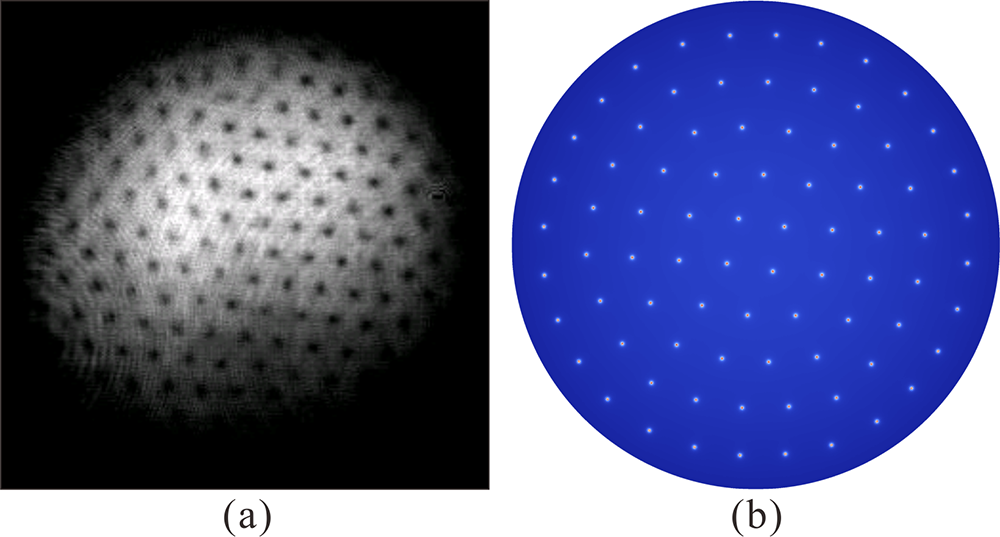}
		\caption{Comparison: (a) an experimental observation of vortex lattice in rotating BEC~\cite{Abo-Shaeer_01} and (b) our simulated vortex lattice result from Fig.~\ref{fig:2d_lattice_result}.
		It is evident that the vortex lattices from the real experiment and our simulation produce very similar spatial structures; each result forms similar triangle-like shapes, within which every vortex is surrounded by a set of vortices that approximately form a hexagonal patten.}
		\label{fig:compare_to_vl}
		\vspace*{-5mm}
	\end{figure}
	
	\begin{figure}[!t]
		\centering
		\includegraphics[width=\columnwidth]{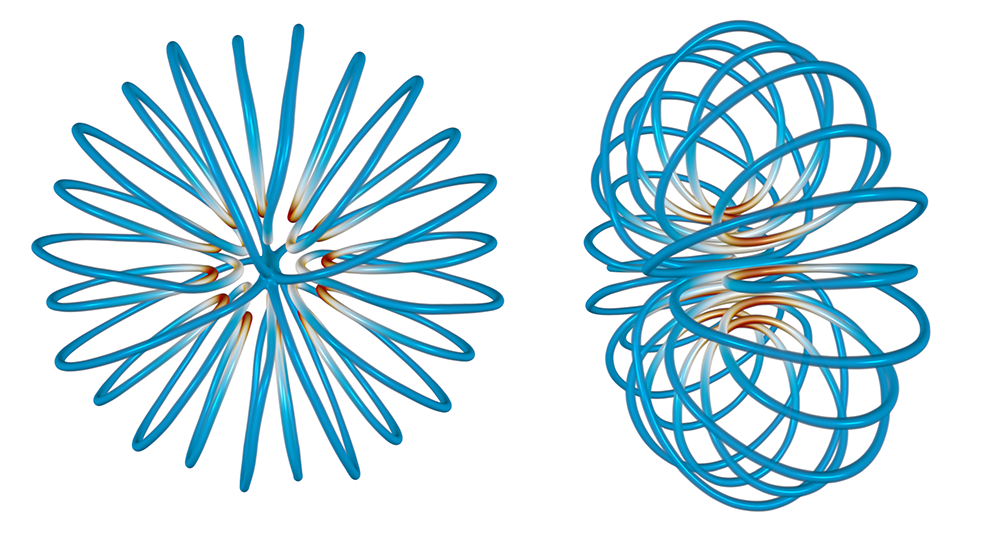}
		\caption{By using a steep Gaussian function with small deviation value and large magnitude as the trapping potential in 3D rotating superfluids, the vortices are completely confined within the simulation grid, where the new vortex ring structures finally emerge.
		Here, we use the x-component of velocity to color-code the surface.}
		\label{fig:3d_vortex_ring}
	\end{figure}
	
	\vspace{0.15cm}
	\noindent
	\textbf{Verification by the Feynman relation.} \
	By increasing the speed of rotation in the system, the number of vortices increases accordingly, see some of the simulation results in Fig.~\ref{fig:2d_lattice_result}.
	Note that we also present the corresponding phase fields, revealing that the vortices always coincide with the phase singularities.
	By automatically detecting the isolated vortex cores in our results with progressively increased rotations, we can obtain a linear relation between the number of vortex cores ($N$) and $\omega \Omega$, see Fig.~\ref{fig:feynman_relation}, showing that our simulation results \rd{approximately} conform to the Feynman relation~\cite{Feynman-55}.
	
	\vspace{0.15cm}
	\noindent
	\textbf{Comparison to real experiments.} \
	Similar vortex lattice structures have already been observed in real experiments with BEC~\cite{Abo-Shaeer_01}, which can be considered as an experimental results of low-speed limit of the superfluid flow.
	Fig.~\ref{fig:compare_to_vl} shows a side-by-side comparison between the real BEC experimental result (Fig.~\ref{fig:compare_to_vl} (a)) and our simulation result (Fig.~\ref{fig:compare_to_vl} (b)).
	It is clear that the spatial distribution of the vortex lattice structure is quite similar between the real experiment and our simulation (see the triangle shapes formed by the vortex cores, and the repeated hexagonal structure of nearby vortex cores for each vortex core).
	However, the two results cannot be made exactly matchable since it is difficult to use the same conditions and system parameters for both experiment and simulation.
	Nevertheless, the close similarity of the vortex lattice distribution gives strong support for the correctness of the simulation as well as the underlying physical model.
	
\begin{figure}[!t]
		\centering
		\includegraphics[width=0.98\columnwidth]{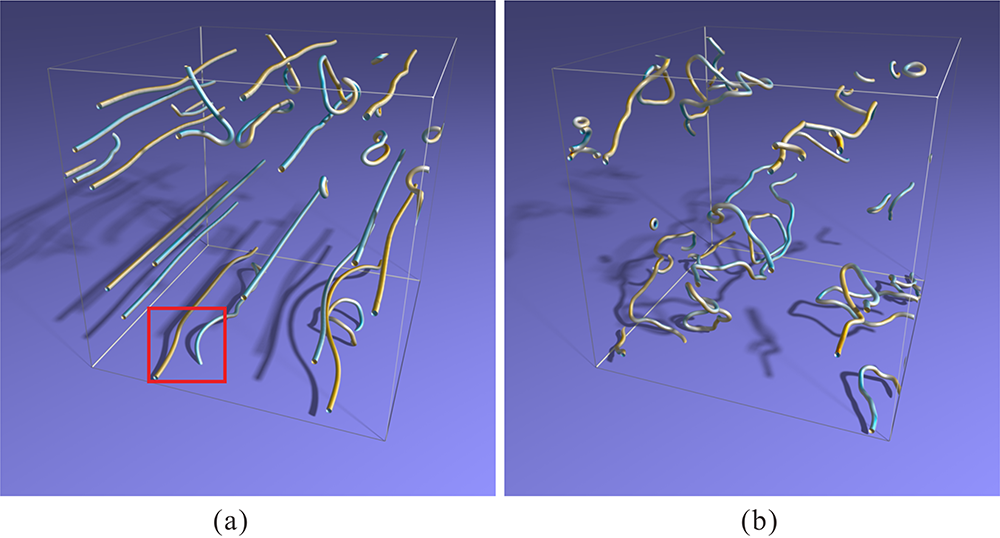}
		\caption{3D global string dynamics:
			(a) vortex strings in an early simulation stage; and
			(b) after a series of reconnections and energy decay.
			The red box \bl{in (a)} marks a pair of anti-parallel vortex lines.}
		\label{fig:3d_global_string}
	\end{figure}

	\vspace{0.15cm}
	\noindent
	\textbf{Vortex ring formation in 3D.}
	In 3D rotating superfluids, the trapping potential is an oblate Gaussian function, such that the vortices are completely confined within the simulation grid, where the vortex ring structures are formed, see Fig.~\ref{fig:3d_vortex_ring} for the result, and Fig.~\ref{fig:teaser} for the snapshots of the related simulation process.
	In Fig.~\ref{fig:teaser}, since the whole magnitude of the velocity is almost the same on an iso-surface, we use the x-component of velocity to color-code the surface, which is helpful to infer the orientation of the vortex lines, e.g., the pairing of anti-parallel vortex lines, as marked by the red box in Fig.~\ref{fig:3d_global_string}.
	
	Furthermore, it is interesting to see that the resulting steady vortex ring structures are nearly symmetric about the center of rotation, but bending to form closed loop structures.
	Note that experimental physicists have been working hard to produce these 3D structures in real experiments, but without success so far.
	\textit{To the best of our knowledge, we are the first in the literature to simulate these vortex ring 3D structures in rotating superfluids}.

	
	\begin{figure}[!t]
		\centering
		\includegraphics[width=0.98\columnwidth]{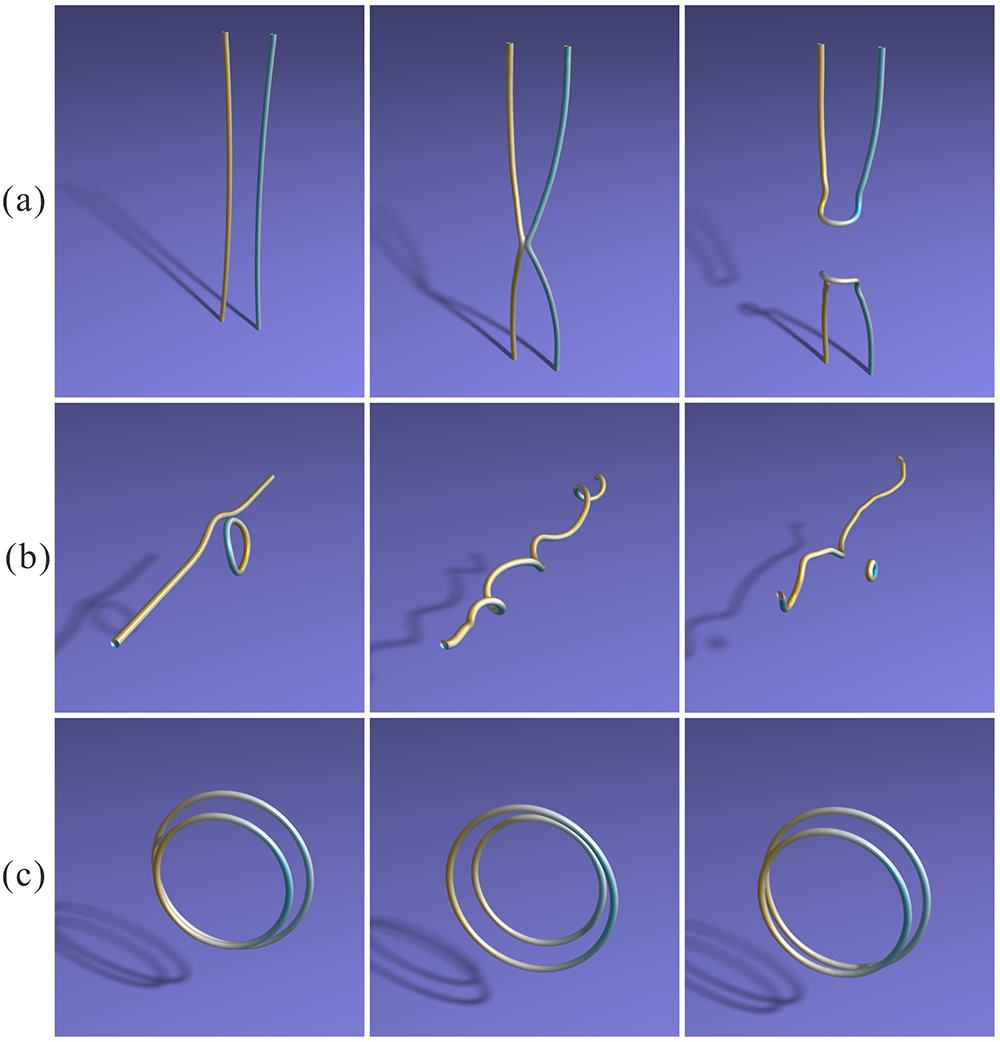}
		\caption{Local vortex dynamics for superfluids:
			(a) local vortex reconnection from two nearby vortex strings;
			(b) Kelvin wave formation by absorbing \bl{energy from} a vortex ring; and
			(c) leapfrogging dynamics in superfluids.
			From left to right: snapshots for the evolution over time.}
		\label{fig:3d_zoomins}
	\end{figure}
	
	\begin{figure}[!t]
		\centering
		\includegraphics[width=0.98\columnwidth]{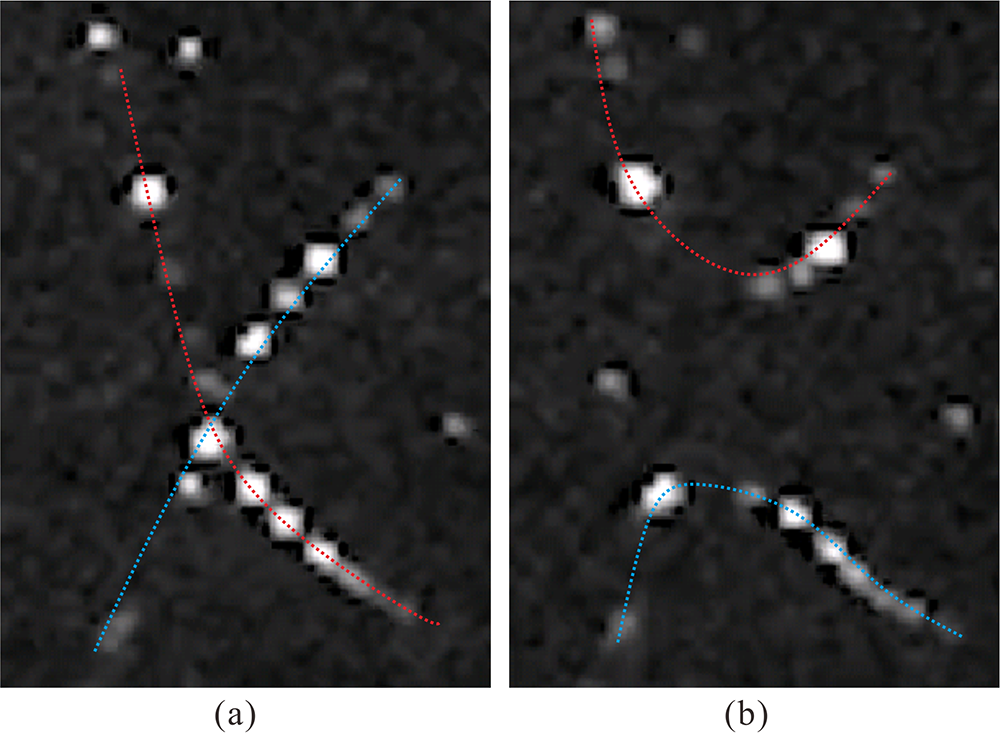}
		\caption{Superfluid liquid Helium vortex reconnection results from experimental physicists~\cite{Dan_14}, showing vortices (a) before and (b) after a reconnection.
			The vortex is visualized by tracer particles.
			We overlay red and blue curves onto the images to highlight individual vortices.
			Note that although the shape of the vortex lines is different from our simulation (Fig~\ref{fig:3d_zoomins} (a)), the spatial topological structure is quite similar, especially for the curved shape after reconnection.
		}
		\label{fig:compare_to_dan}
	\end{figure}


	\subsection{Unsteady Vortex Dynamics}
	\label{sec:unsteady_vortex_dyanmics}
	
	Superfluids without constant rotation will not have steady solutions.
	The vortices will always change their forms over time, resulting in very complicated vortex structures.
	Here, we study the properties for both global and local vortex dynamics, where we set the system parameters as: domain size $L = 200$, $\lambda = 1$, $F_0 = 1$, $m_0 = 1$, time step $\Delta t= 0.1$, and resolution is $512^3$.
	In particular, we study two important local vortex dynamics that form the basis of the global vortex dynamics: the {\em vortex-vortex reconnection} and {\em the Kelvin wave formation and transmission}.
	In addition, we also study {\em vortex tangle} and the infinite {\em leapfrogging phenomenon}, which is a very special vortex dynamics.
	
	\vspace{0.15cm}
	\noindent
	\textbf{Local vortex reconnection.} \
	Although global vortex dynamics (including vortex tangle) in superfluids is rather complicated, it can be decomposed into several characteristic basic local dynamics.
	One of them is the local vortex reconnection illustrated in Fig.~\ref{fig:3d_zoomins}(a).
	From the figure, it is clear that when two vortex strings are close to each other at certain velocity, they may collide.
	The result of the collision will be a re-organization of the vortex string topology by combining the old vortex strings and reproducing the new ones.
	Such reconnection phenomenon has been observed experimentally in superfluid liquid Helium~\cite{Dan_14}, see results in Fig.~\ref{fig:compare_to_dan}; we can see that our simulated and visualized results can produce similar dynamics, as compared to Fig.~\ref{fig:3d_zoomins}(a).

	Note that like the comparison for vortex lattice formation, it is difficult to have exactly the same side-by-side comparison for the vortex reconnection between experiment and simulation since it is hard to have the same system conditions and parameter settings.
	However, we can still make some comparison and find out the similarity between their spatial topological structures, which gives strong evidence that the model equation we use and simulation methods we develop conform to the real physics.

	\vspace{0.15cm}
	\noindent
	\textbf{Kelvin wave.} \
	Another important basic local dynamics is the Kelvin wave formation and transmission, which can be easily triggered and observed when vortex structures reconnect.
	Here, we demonstrate this phenomenon with Fig.~\ref{fig:3d_zoomins}(b), where a vortex ring approaches a vortex line.
	After colliding with the vortex line, the vortex ring will be absorbed and a proportion of its energy will be sent out in the form of helical Kelvin waves from the collision points to the two sides of the vortex line.
	Finally, due to the periodic boundary condition, the waves will collide with its two wavefronts along the vortex line and a new vortex ring will be ``split out'' from the vortex line but with smaller ring size, revealing that Kelvin wave is actually a driving force for successive reconnection events in superfluids.

	\vspace{0.15cm}
	\noindent
	\textbf{Global vortex dynamics.} \
	We produce global vortex dynamics in a non-rotating system using initial vortex lines taken from intermediate solutions of a 3D rotating superfluid simulation.
	Without rotation, the vortices will not remain steady and will evolve freely over time, see Fig.~\ref{fig:3d_global_string}.
	Long vortex lines will be further split by vortex reconnections into shorter vortex lines, and as time proceeds, merging process will also happen to reconnect shorter vortex lines into longer ones.
	Such split and merge may exist simultaneously in the domain, and the whole process iterates over time to produce more complicated vortex structures with irregular topology, see Fig.~\ref{fig:3d_global_string}(b).
	
\begin{figure}[!t]
		\centering
		\includegraphics[width=0.98\columnwidth]{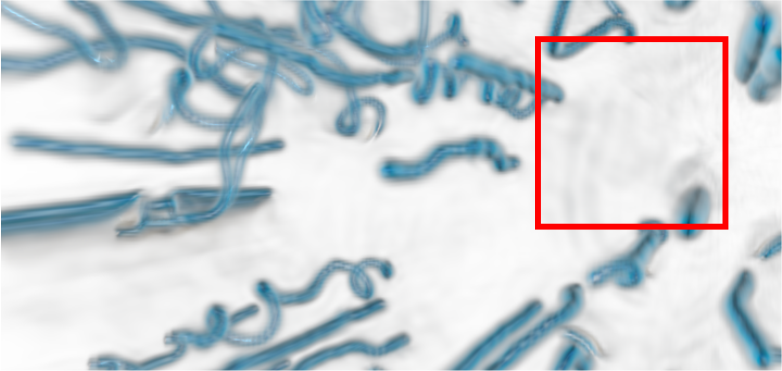}
		\caption{\bl{By applying direct volume rendering,} compression waves can be observed in the field magnitude.
			These waves, e.g., see the red boxed region, are energy releases from the vortex reconnections, and they accumulate \bl{by the transmission through} the periodic boundary.
		}
		\label{fig:3d_waves}
	\end{figure}

	It is important to note that during each vortex reconnection in global vortex dynamics, some energy will be released out from the vortices in the form of waves, see Fig.~\ref{fig:3d_waves} for a corresponding zoom-in view produced by using direct volume rendering.
	It is such energy release that decays the vortices in the whole system, see again Fig.~\ref{fig:3d_global_string} (a \& b).
	Hence, certain amount of energy is needed to add into the system to maintain the overall vortex structures during the reconnections.
	
	\begin{figure*}[!t]
		\centering
		\includegraphics[width=0.98\textwidth]{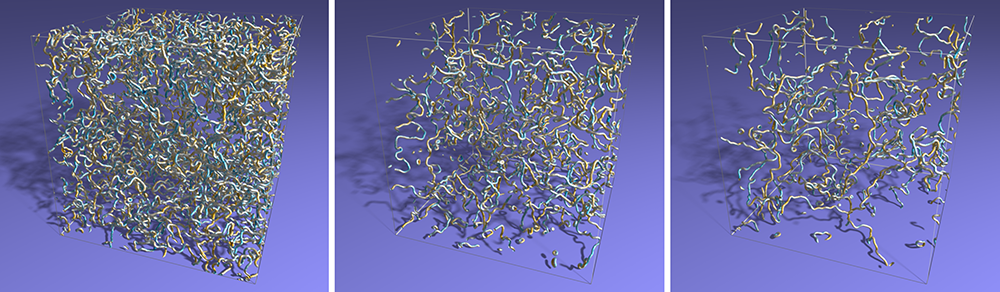}
		\caption{Snapshots showing the decaying process of the 3D vortex tangle.
		Note that the energy release during the massive reconnection processes dacays the overall vortices over time.}
		\label{fig:3d_tangle}
	\end{figure*}
	
	\vspace{0.15cm}
	\noindent
	\textbf{Vortex tangle.} \
	Using a few vortex lines taken from a 3D rotating superfluid cannot provide sufficiently dense vortex strings to achieve the state of vortex tangle.
	As described earlier in Section~\ref{sec:3d_simu}, if we inject vortex rings randomly from the domain boundary, we can effectively create dense amount of vortex reconnections and generate the state of vortex tangle, see Fig.~\ref{fig:3d_tangle}.
	Note that the vortex tangle is decaying quantum turbulence, where the vortices are reduced by energy release during the massive reconnection processes.
	
	\vspace{0.15cm}
	\noindent
	\textbf{Leapfrogging phenomena.} \
	Lastly, we study one interesting vortex dynamics phenomenon called ``leapfrogging.''
	Leapfrogging phenomena have been observed in classical fluids, and can also be simulated~\cite{Riley_93}.
	However, it has not yet been observed in superfluids.
	By initializing $\Phi$ with two coplanar vortex rings with the same orientation, we can simulate stable leapfrogging phenomena, see Fig.~\ref{fig:3d_zoomins}(c).
	\textit{Please refer to the supplementary video for the corresponding animated simulations and visualizations.}
	
	Our results appear similar to the leapfrogging in classical fluids, where the two rings alternatively move forward through the other with varying radius.
	However, due to vorticity diffusion and turbulence, leapfrogging in classical fluids can be destroyed easily, \bl{and cannot last for a long time}.
	In the case of superfluids, due to zero viscosity, vortices do not merge or dissipate, thus leading to more stable and long-lasting leapfrogging phenomena.
	If we consider periodic boundaries in the spatial domain, a two-vortex-ring leapfrogging system can proceed indefinitely.

	
	\subsection{Implementation and Discussion}
	
	\noindent
	\textbf{Implementation details.} \
	We implemented and ran our superfluid simulation and vortex structure visualization on a workstation with an Intel Xeon E5-2630 v3 CPU, 32GB system memory and an NVIDIA GTX TITAN X GPU.
	The solvers and renderers are all written in C++ and CUDA. 
	The simulations and visualizations are accelerated by the GPU due to its nice parallelism feature.
	
	The average time for producing the simulations is 9 to 10 seconds for 10000 iterations in 2D with a grid resolution of $256 (r) \times 1024 (\theta)$ in polar coordinates. 
	The average time is around 0.1 second per iteration in 3D with a grid resolution of $512^3$ in Cartesian coordinates.
	The 2D renderings are straightforward, but it takes even more time than simulation because we use Matlab for visualization, and call our CUDA kernels through the MEX interface of Matlab. 
	For 3D simulation, the \bl{vortex identification} is very fast; it takes a few seconds to \bl{obtain} the circulation field and volumetrically render the structure at a grid resolution of $512^3$, such that the 3D simulation can be previewed without much performance drop. 
	For higher quality, we build our offline volume raytracer for the \bl{generated vortex tube} iso-surfaces, where the rendering cost is as high as a few minutes per frame.
	
	\vspace{0.1cm}
	\noindent
	\textbf{Initialization.}
	In general, we require the initialization to contain large enough circulation, but we can use different distributions of the field, which does not influence the final vortex phenomena, for no matter steady or unsteady superfluid vortex dynamics.
	For steady solutions on rotating superfluids, the solution always converges to the steady \rd{vortex lattice (or vortex ring in 3D) structure}, while for unsteady solutions, we are interested in either local dynamics (such as reconnections and Kelvin waves) or vortex tangle.
	In either case, \rd{although some initial conditions are
		artificial, the underlying motions are strictly under the corresponding physical law by the model, which can result in believable phenomena that could be observed in practice, and our method can produce some phenomena that can be truly observed in real experiments.}
	
	\vspace{0.1cm}
	\noindent
	\textbf{Accuracy analysis.}
	Since we use high-order numerical methods in both 2D and 3D, the original discretization of the model equation is accurate enough for visualization purpose， especially if we use high resolution grid.
	However, due to strong waves which are eventually of high frequency, we employ a temporal filtering method to make the simulation data clear enough for vortex identification.
	This is based on an assumption that vortices and compression waves are completely separable in time frequency.
	However, if there are vortices that vary as fast as waves, these vortices will be removed, which introduces a certain degree of uncertainty in our final results.
	However, by our experiments and observations, such phenomenon is very rare \rd{in most cases}, as the waves are finally of extremely high frequency everywhere.
		
	On the other hand, during vortex visualization, we employ circulation with a loop size of two grid cells to identify the vortex cores.
	Here, we also make an assumption that all the vortex cores are of the same scale, which is true for superfluids and in contrast to classical fluids where vortices can have varying scales.
	With circulation-based approach, we are uncertain about the exact location of the vortex core, where some small-scale (scale smaller than two grid cells) variation of the vortex structure is missing.
	However, we are sure that the vortex tubes we generated and visualized can truly reflect the large scale (scale larger than two grid cells) structure of the vortex variations since the vortex cores are completely included inside the circulation loops.
		
	Since large-scale structures are more meaningful than small fluctuations for most of the cases, the visualization we generated can be useful enough for scientific comparison and visual analysis, as done in Sections~\ref{sec:steady_vortex_dyanmics} \& \ref{sec:unsteady_vortex_dyanmics}.
	For more accurate analysis involving smaller-scale structures, we only need to increase the grid resolution, and our vortex identification and visualization methods can automatically obtain smaller scale structures since the circulation loop is always of two grid cell size.
	
	\vspace{0.1cm}
	\noindent
	\textbf{Feedback from physicists.} \
	Our visualization results have been given to two research groups of superfluid scientists for feedback.
	One of such groups are our collaborators at Nanyang Technological University, led by Professor Emeritus Kerson Huang from MIT, who are all theoretical physicists.
	They felt that our visualizations and animations have more intuitively helped them observe the superfluid structure from the physical model, which has given them deeper understanding, especially in 3D, and find out more interesting and useful properties for superfluid vortices, as evidenced by a recent
	paper published by the group~\cite{PRD_14}.
	Another group of scientists are from the research group led by Professor Daniel Lathrop at University of Maryland, who are all experimental scientists on superfluids.
	They gave us very positive feedback for our simulation and visualization results, with a particular interest on Kelvin wave simulation and visualization.
	With our results, both groups of scientists can easily analyze the steady and unsteady behavior of superfluid vortices on a computer, and even discover new vortex structures more intuitively by trying different sets of system parameters. 
	This can provide some guidance in experimental observations, \rd{which indicates the significance of our simulation and visualization work to help support research in fundamental science}.

	\vspace{0.1cm}
	\noindent
	\textbf{Limitations.}
	Since the vortex cores exist as singularities in the phase field of the NLKG equation and are infinitesimally small, we cannot exactly locate them and visualize them with thin-enough structures at proper grid resolution.
	In addition, our visualization of superfluid vortices is done by extracting iso-surfaces of a smoothed circulation field, and the vortex tube size is larger than the grid spacing. 
	Thus, very dense vortex tangles cannot be visualized clearly.
	Increasing the grid resolution can alleviate such problem and make the tube thinner to better reflect the vortex core geometry, but at the expense of more computing resource and time.
	Finally, the Laplacian smoothing on the circulation field may also remove some vortices given the limited grid resolution.
	
	\section{Conclusion}
	\label{sec:conclusion}
	
	In this work, we aim for high-quality superfluid visualization and \rd{to help our collaborators visually study superfluid vortices.}
	To achieve this goal, we solve the underlying nonlinear Klein-Gordon equation to prepare proper datasets for visualization, and develop an effective vortex core identification method based on a particular loop selection model and an orthogonal-plane strategy, which are local and efficient for parallel implementation on GPU.
	Visual analysis of superfluid vortex structures is conducted with domain experts by considering both steady and unsteady scenarios, where we observe vortex lattice and ring structures in 2D \& 3D, approximate the Feynman relation in 2D, and produce local vortex reconnections, Kelvin waves, vortex tangles and leapfrogging phenomena in 3D.
	These results helped physicists to better understand the stationary as well as dynamic properties of superfluid vortices.
	For verification purpose, we compared our visualization with experimental results and the Feynman relation.
	
	As a future work, we would like to explore the use of adaptive grids to produce higher resolution vortex core regions to more accurately locate vortex cores and reveal finer-scale structures.
	Such result will be useful for comparative studies with real experiments.
	Moreover, we plan to explore multi-scale visualization, where we visualize small-scale vortex core structures on top of large-scale structures for studying the relation and interaction between vortex core structures across different scales.
	
	\section*{Acknowledgment}
	The authors would like to thank all anonymous reviewers for their constructive comments.
	This work is supported by National Science Foundation of China (NSFC) - Outstanding Youth Foundation (Grant No. 61502305), ShanghaiTech University startup funding, Guangdong high-level personnel of special support program (Grant No. 2016TQ03X319), and the CUHK strategic recruitment fund and direct grant (4055061).

	\bibliographystyle{IEEEtran}


\end{document}